\documentclass{article}
\usepackage{arxiv}

\usepackage{amsmath,amsfonts,bm}

\def\Figref#1{Figure~\ref{#1}}

\def\eqref#1{equation~\ref{#1}}
\def\Eqref#1{Equation~\ref{#1}}

\def\1{\bm{1}}

\DeclareMathAlphabet{\mathsfit}{\encodingdefault}{\sfdefault}{m}{sl}
\SetMathAlphabet{\mathsfit}{bold}{\encodingdefault}{\sfdefault}{bx}{n}

\usepackage{graphicx}
\usepackage{xcolor}
\usepackage{wrapfig}
\usepackage{hyperref}
\usepackage{url}
\usepackage{enumitem}
\usepackage{booktabs}
\usepackage{xcolor}
\usepackage{colortbl}
\usepackage{amssymb}
\usepackage{natbib}
\newcommand{\cmark}{\ensuremath{\checkmark}}
\newcommand{\xmark}{\ensuremath{\times}}
\definecolor{trainblue}{HTML}{3478C0}
\definecolor{heldpurple}{HTML}{8B58B8}
\definecolor{SectionBlue}{RGB}{245,248,252}

\newcommand{\resultvalue}[2]{%
  #1\,\textcolor{black!50}{\scalebox{0.78}{$\pm\,#2$}}%
}

\newcommand{\firstresult}[2]{%
  \textbf{#1}$^{\ast}$\,
  \textcolor{black!50}{\scalebox{0.78}{$\pm\,#2$}}%
}

\newcommand{\secondresult}[2]{%
  #1$^{\dagger}$\,
  \textcolor{black!50}{\scalebox{0.78}{$\pm\,#2$}}%
}

\newcommand{\thirdresult}[2]{%
  #1$^{\ddagger}$\,
  \textcolor{black!50}{\scalebox{0.78}{$\pm\,#2$}}%
}

\newcommand{\sectionheader}[1]{%
  \rowcolor{SectionBlue}%
  \multicolumn{10}{@{}l}{\textbf{#1}}\\%
}

\title{BEVLight: Generalizable Traffic Signal Control from UAV Bird's-Eye Views}

\author{
\textbf{Maonan Wang}$^{1,2}$\thanks{Equal contribution.},
\textbf{Yuxin Cai}$^{3}$\footnotemark[1],
\textbf{Junming Liu}$^{2}$,
\textbf{Zhengyan Huang}$^{2}$, \\
\textbf{Yi Yu}$^{2}$,
\textbf{Ding Wang}$^{2}$,
\textbf{Yifeng Zhang}$^{4}$,
\textbf{Yirong Chen}$^{2}$\thanks{Corresponding author.} \\
$^{1}$ The Chinese University of Hong Kong, Shenzhen \quad
$^{2}$ Shanghai AI Lab \\
$^{3}$ Nanyang Technological University \quad
$^{4}$ National University of Singapore
}

\date{}

\begin{document}
\maketitle

\begin{abstract}
Learning-based traffic signal control (TSC) has achieved promising performance, but most methods rely on structured traffic states that must be estimated from sensor observations in real deployments. Visual TSC learns signal decisions directly from images, yet existing methods depend on fixed roadside cameras and remain tied to specific intersection configurations and visual domains. UAVs offer on-demand bird's-eye-view (BEV) observations that jointly capture intersection topology and traffic dynamics, making them a promising sensing platform for visual TSC. We present BEVLight, a generalizable TSC framework that directly maps BEV observations to signal actions. BEVLight grounds visual features onto individual lanes and composes them along the junction topology into movement and phase representations, allowing a shared policy to score candidate phases across heterogeneous intersection layouts and signal plans. To train the policy, we develop a web-based virtual teleoperation platform and collect over 5K human control decisions together with their underlying traffic states. Simulator-provided traffic states first supervise lane- and movement-level representations, after which human phase selections and rankings supervise the phase scorer. Re-rendering the recorded trajectories under diverse appearances preserves both forms of supervision and encourages robustness to visual changes. Closed-loop experiments demonstrate zero-shot generalization across traffic demands, signal plans, and intersections: BEVLight reduces average waiting time by 7--9\% relative to perception-to-control visual baselines and matches or outperforms structured-state RL controllers that receive ground-truth traffic states, achieving the lowest waiting time in 12 of 16 scenarios. An offline study further shows that BEVLight transfers to real UAV observations without target-domain adaptation.

\end{abstract}

\section{Introduction}
\label{sec:introduction}

Traffic signal control (TSC) requires decisions that adapt to evolving traffic conditions across intersections with different geometries. Recent controllers, from RL policies to LLM agents, have shown promising results, but most operate on structured traffic states such as vehicle counts, queue lengths, and vehicle positions~\citep{wei2018intellilight, wei2019presslight, chen2020toward, wang2024unitsa, lai2025llmlight, yuan2026collmlight}. These states are readily available in simulation but must be estimated in practice, requiring additional perception pipelines whose errors propagate into control.

\begin{figure}[t]
  \centering
  \includegraphics[width=\textwidth]{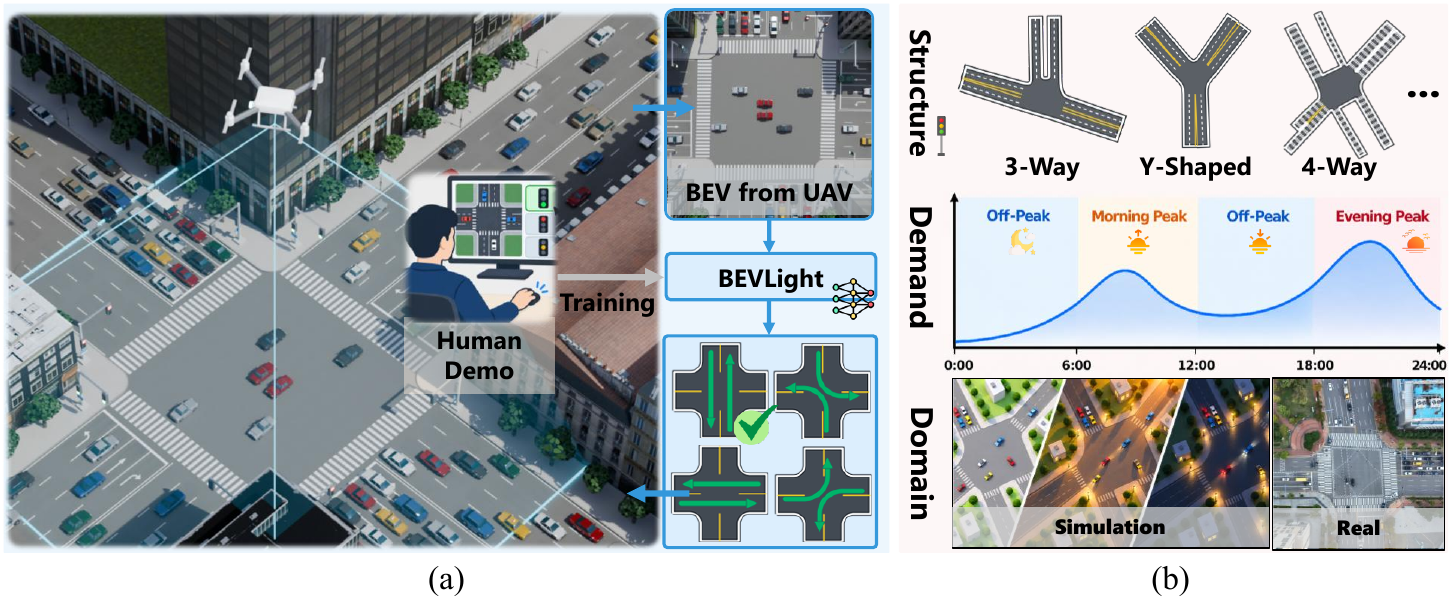}
  \caption{(a) UAV-based visual TSC: BEVLight maps BEV observations to signal phases and is trained from human demonstrations that select and rank feasible phases. (b) The three generalization dimensions considered in this work: structure, demand, and domain.}
  \label{fig:teaser}
\end{figure}

Visual TSC instead learns signal decisions directly from images~\citep{garg2022vision,chen2024syntrac,wang2025vlmlight}, removing the need for explicit traffic-state estimation. Existing approaches, however, typically consider a fixed or limited set of intersections and signal plans, and obtain observations from roadside cameras that require site-specific installation and provide fixed coverage. UAVs offer an alternative: they can be deployed on demand without permanent roadside infrastructure~\citep{menouar2017uav,fonod2025advanced}, and their elevated viewpoints provide bird's-eye-view (BEV) observations that jointly capture intersection topology and traffic conditions, as illustrated in \Figref{fig:teaser}(a).

A key challenge is therefore to learn a single UAV-based visual TSC policy that generalizes across deployment conditions. As illustrated in \Figref{fig:teaser}(b), we consider three dimensions of generalization. Structure covers junction geometry and signal configuration, from different signal plans at the same intersection to unseen intersections with different lane topologies and feasible phase sets; across these, the same action index can correspond to different movements, so the policy cannot rely on a fixed action layout. Demand reflects variations in traffic flow that produce different congestion and queue patterns at the same intersection. Domain captures appearance changes from weather, illumination, and the gap between simulated and real UAV imagery. Simulation can expose a policy to combinations of these variations that are impractical to collect at real intersections, provided the policy is not tied to any structure, demand, or appearance.

To address these challenges, we propose BEVLight, a generalizable UAV-based visual TSC framework that maps BEV observations directly to signal actions. BEVLight grounds visual features onto individual lanes and composes them along the junction topology into phase representations, so that a shared scorer ranks candidate phases across layouts and signal plans. Training such a scorer requires supervision on how competing phases should be prioritized, not only which phase is chosen. We therefore learn the policy from human demonstrations that rank all feasible phases at each decision, a preference structure that neither scalar RL rewards nor single-action labels provide. To collect them, we develop a virtual teleoperation platform in which participants control simulated intersections from a BEV view, selecting and ranking feasible phases while the platform records the traffic trajectories. Simulator-provided traffic states first supervise lane- and movement-level representations, after which human selections and rankings supervise the phase scorer. Re-rendering the recorded trajectories under diverse appearances preserves both forms of supervision and encourages appearance-invariant representations.

The contributions of our work are summarized as follows:
\begin{itemize}[leftmargin=*]
    \item We propose BEVLight, a topology-grounded framework for TSC from UAV BEV observations. By composing visual lane features into movement and phase representations, a single policy handles different intersection layouts and signal plans without retraining.
    \item We learn signal decisions from human demonstrations that rank all feasible phases, providing preference supervision that a shared phase scorer exploits across signal plans with different phase sets. A virtual teleoperation platform supports collecting over 5K such ranked decisions and re-rendering them under varied visual appearances.
    \item We demonstrate zero-shot generalization across traffic demands, signal plans, and intersections. BEVLight reduces average waiting time by 7--9\% relative to perception-to-control visual baselines, and matches or outperforms structured-state RL methods that receive ground-truth traffic states, achieving the lowest waiting time in 12 of 16 scenarios. An offline study further shows transfer to real UAV videos without target-domain adaptation.
    
\end{itemize}

\section{Preliminaries and Problem Formulation}
\label{sec:preliminaries}

\subsection{Traffic Signal Control Preliminaries}
\label{sec:tsc_preliminaries}

\begin{wrapfigure}[17]{r}{0.55\textwidth}
    \centering
    \includegraphics[width=\linewidth]{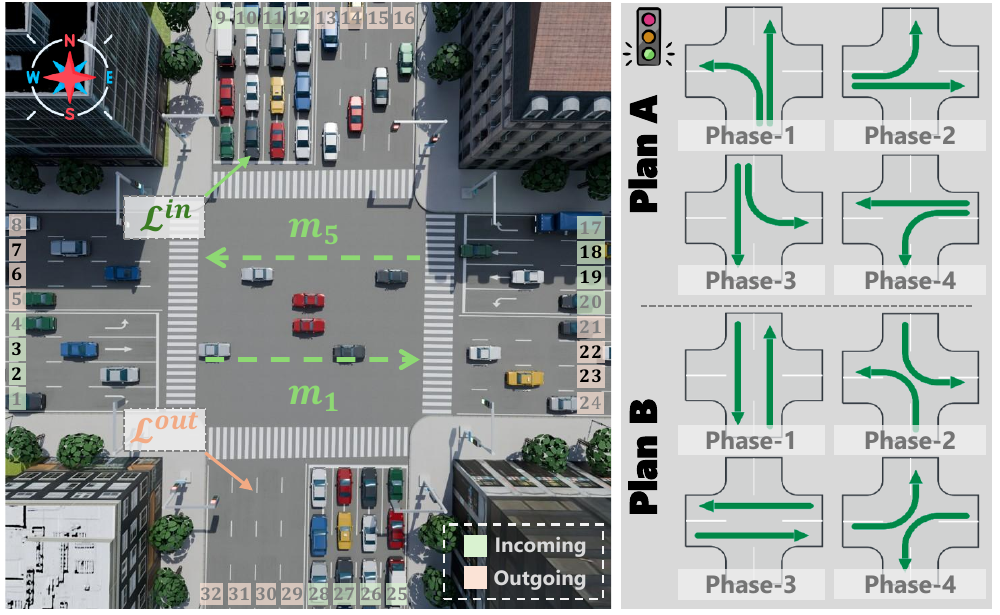}
    \caption{Illustration of a four-way intersection under two phase configurations, each comprising four traffic signal phases.}
    \label{fig:tsc_preliminaries}
\end{wrapfigure}

\Figref{fig:tsc_preliminaries} illustrates a four-way intersection under two phase configurations.


\paragraph{Lane.} Incoming lanes $\mathcal{L}^{\mathrm{in}}$ carry vehicles toward the intersection and outgoing lanes $\mathcal{L}^{\mathrm{out}}$ carry them away, with $\mathcal{L}=\mathcal{L}^{\mathrm{in}}\cup\mathcal{L}^{\mathrm{out}}$; in \Figref{fig:tsc_preliminaries}, $\ell_1$--$\ell_5$ are incoming and $\ell_{21}$--$\ell_{24}$ are outgoing.

\paragraph{Movement.} A movement $m\subseteq\mathcal{L}^{\mathrm{in}}\times\mathcal{L}^{\mathrm{out}}$ groups admissible lane connections that serve the same directional flow and share signal control, e.g., the opposing through movements $m_1=\{(\ell_2,\ell_{23}),(\ell_3,\ell_{22})\}$ and $m_5=\{(\ell_{19},\ell_6),(\ell_{18},\ell_7)\}$. We denote the movement set by $\mathcal{M}$.


\paragraph{Phase.} A phase $p\subseteq\mathcal{M}$ contains non-conflicting movements that receive the right of way simultaneously, and a phase configuration defines the feasible phase set $\mathcal{P}=\{p_1,\ldots,p_K\}$. Plan~A and Plan~B in \Figref{fig:tsc_preliminaries} group the same movements differently: Plan~A releases $\{m_1,m_5\}$ as Phase-3, whereas Plan~B pairs them with other movements.

\subsection{UAV Bird's-Eye-View TSC}
\label{sec:uav_bev_tsc}

We characterize a junction by its configuration
\begin{equation}
    \mathcal{C}=(\mathbf{M}_{\mathcal{L}},\mathcal{M},\mathcal{P}),
    \qquad
    \mathbf{M}_{\mathcal{L}}
    =\{\mathbf{M}_\ell:\ell\in\mathcal{L}\},
    \label{eq:configuration}
\end{equation}
where $\mathbf{M}_\ell\in\{0,1\}^{H_{\mathrm{BEV}}\times W_{\mathrm{BEV}}}$ is the binary BEV mask of lane $\ell$, $\mathcal{M}$ is the movement set, and $\mathcal{P}$ is the feasible phase set of the current signal plan. At decision step $t$, the controller receives
\begin{equation}
    o_t=\left(\mathbf{I}_t^{\mathrm{BEV}},\mathcal{C},k_t,\tau_t\right),
    \qquad
    \mathbf{I}_t^{\mathrm{BEV}}
    =\left(I_{t-H+1}^{\mathrm{BEV}},\ldots,I_t^{\mathrm{BEV}}\right),
    \label{eq:observation}
\end{equation}
where $\mathbf{I}_t^{\mathrm{BEV}}$ collects the latest $H$ UAV images, $k_t\in\{1,\ldots,K\}$ indexes the current phase $p_{k_t}$, and $\tau_t\geq0$ is its elapsed duration. The policy selects a feasible target phase $a_t=\pi_\theta(o_t)\in\mathcal{P}$. Our goal is a single policy $\pi_\theta$ that generalizes across junction layouts, signal plans, and traffic demands, and transfers from simulated BEV sequences to real UAV observations.

\section{Method}
\label{sec:method}

\begin{figure}[t]
  \centering
  \includegraphics[width=\textwidth]{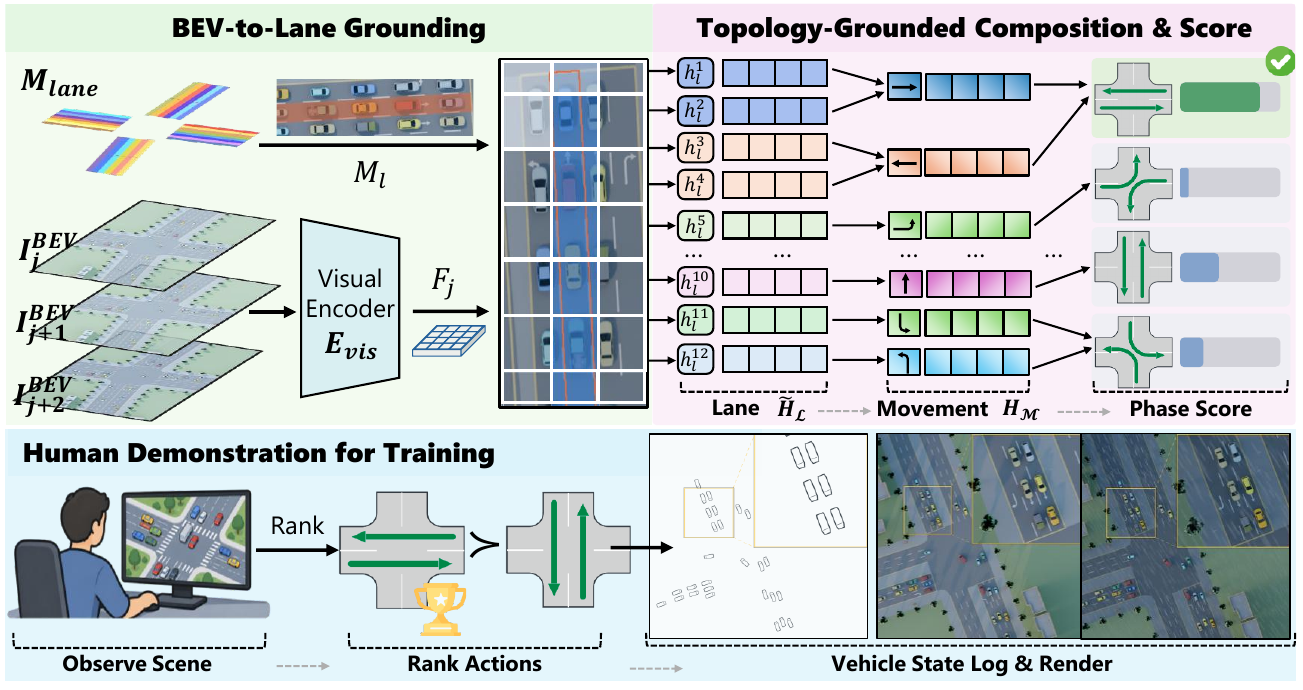}
  \caption{Overview of BEVLight. Visual patch features are grounded onto individual lanes using lane masks with content-adaptive pooling (BEV-to-Lane Grounding). Lane representations are composed along the junction topology into movement and phase representations, and a shared scorer ranks candidate phases (Topology-Grounded Composition \& Score). Simulator-provided traffic states supervise lane and movement representations, while human phase selections and rankings supervise the scorer; recorded trajectories are re-rendered under diverse appearances with unchanged supervision (Human Demonstration for Training).}
  \label{fig:framework}
\end{figure}

BEVLight learns a shared visual policy across junction geometries and signal configurations. Since BEV observations cannot directly map to junction-specific action indices, BEVLight grounds visual features onto lanes and uses traffic topology to construct movement and phase representations with consistent semantics. As illustrated in \Figref{fig:framework}, the framework comprises BEV-to-lane grounding, topology-grounded composition and scoring, and human-supervised policy learning. The policy is trained from human traffic-control demonstrations together with traffic-state supervision encouraging visual representations to preserve control-relevant information.

\subsection{Topology-Grounded BEV-to-Phase Representation}
\label{sec:representation}

A consistent mapping from BEV observations to signal actions must account for junction-specific geometries and signal plans, under which the same action index can correspond to different movements (\Figref{fig:tsc_preliminaries}). BEVLight resolves this ambiguity by grounding visual features onto individual lanes and following the junction topology to progressively compose lane representations into movement and phase representations.

\paragraph{BEV-to-Lane Grounding.}
For each $I_j^{\mathrm{BEV}}$ in $\mathbf{I}_t^{\mathrm{BEV}}$, the visual encoder produces $U$ patch features,
\begin{equation}
    \mathbf{F}_j
    =
    E_{\mathrm{vis}}(I_j^{\mathrm{BEV}})
    =
    [\mathbf{f}_1^j,\ldots,\mathbf{f}_U^j],
\end{equation}
where $\mathbf{f}_u^j$ denotes the feature of patch $u$.
To associate these patches with lane $\ell$, we derive two geometric attributes from its lane mask $\mathbf{M}_\ell$: the coverage coefficient $c_{\ell,u}\in[0,1]$, which measures the fraction of patch $u$ covered by the lane, and the normalized longitudinal offset $o_{\ell,u}=d_{\ell,u}/L_\ell$, where $d_{\ell,u}$ is the distance from patch $u$ to the stop line along the lane and $L_\ell$ is the lane length.

Rather than uniformly pooling the patches overlapping a lane, BEVLight assigns content-adaptive weights by combining a mask-derived geometric prior with a learned correction:
\begin{equation}
    a_{\ell,u}^{j}
    =
    \underbrace{
    \log(c_{\ell,u}+\epsilon)
    }_{\text{geometric mask prior}}
    +
    \underbrace{
    \operatorname{MLP}_{\mathrm{pool}}
    \left(
    \operatorname{concat}
    [\mathbf{f}_u^j,c_{\ell,u},o_{\ell,u}]
    \right)
    }_{\text{content-dependent correction}},
    \label{eq:lane_pool_score}
\end{equation}
where $\epsilon>0$ ensures numerical stability. A softmax over the patches overlapping $\mathbf{M}_\ell$ gives $\alpha_{\ell,u}^{j} = \operatorname{Softmax}_{u:c_{\ell,u}>0}(a_{\ell,u}^{j})$. The frame-level lane token is then
\begin{equation}
    \mathbf{v}_{\ell}^{j}
    =
    \sum_{u}
    \alpha_{\ell,u}^{j}
    \left(
    \mathbf{f}_u^j
    +
    \operatorname{Fourier}(o_{\ell,u})
    \right).
    \label{eq:lane_token}
\end{equation}
The longitudinal offset, therefore, influences both the adaptive pooling weights and the positional information retained in the resulting lane representation.

Given the lane tokens from consecutive BEV frames, we further model their short-term temporal evolution and interactions across lanes, producing contextualized lane representations $\widetilde{\mathbf{H}}_{\mathcal L} = [\widetilde{\mathbf h}_{\ell}^{\mathrm L}]_{\ell\in\mathcal L}$. The complete temporal and inter-lane modeling procedure is detailed in Appendix~\ref{app:experiment_settings}.

\paragraph{Topology-Grounded Composition.}
We next propagate the lane representations through the junction topology. Three fixed binary incidence matrices derived from the junction configuration $\mathcal{C}$, $\mathbf{B}_{\mathcal L\rightarrow\mathcal M}^{\mathrm{in}}$, $\mathbf{B}_{\mathcal L\rightarrow\mathcal M}^{\mathrm{out}}$, and $\mathbf{B}_{\mathcal M\rightarrow\mathcal P}$, encode incoming lane-to-movement, outgoing lane-to-movement, and movement-to-phase relations, respectively. For each movement, we first aggregate its incoming and outgoing lane context:
\begin{equation}
    \mathbf{H}_{\mathcal M}^{\mathrm{in}}
    =
    \mathbf{B}_{\mathcal L\rightarrow\mathcal M}^{\mathrm{in}}
    \widetilde{\mathbf{H}}_{\mathcal L},
    \qquad
    \mathbf{H}_{\mathcal M}^{\mathrm{out}}
    =
    \mathbf{B}_{\mathcal L\rightarrow\mathcal M}^{\mathrm{out}}
    \widetilde{\mathbf{H}}_{\mathcal L}.
    \label{eq:movement_aggregation}
\end{equation}
For movement $m$, a shared encoder combines the incoming and outgoing summaries together with their difference,
\begin{equation}
    \bar{\mathbf{h}}_{m}^{\mathrm M}
    =
    \operatorname{MLP}_{M}
    \left(
    \operatorname{concat}
    [
    \mathbf{h}_{m}^{\mathrm{M,in}},
    \mathbf{h}_{m}^{\mathrm{M,out}},
    \mathbf{h}_{m}^{\mathrm{M,in}}
    -
    \mathbf{h}_{m}^{\mathrm{M,out}}
    ]
    \right),
    \label{eq:movement_token}
\end{equation}
where the incoming--outgoing difference provides a pressure-like cue for traffic imbalance. We further contextualize the movement representations via self-attention across movements to obtain $\mathbf{h}_{m}^{\mathrm M}$ (Appendix~\ref{app:architecture}).

We then compose the movements associated with each feasible phase according to $\mathbf{B}_{\mathcal M\rightarrow\mathcal P}$.
For phase $p$, let
$\mathbf{H}_{p}=[\mathbf{h}_{m}^{\mathrm M}]_{m\in p}$.
A shared learnable phase query summarizes the relative composition of its constituent movements,
\begin{equation}
    \mathbf{c}_{p}
    =
    \operatorname{CrossAttn}
    \left(
    \mathbf{q}_{P},
    \mathbf{H}_{p},
    \mathbf{H}_{p}
    \right).
    \label{eq:phase_attention}
\end{equation}
Since normalized attention primarily captures relative composition, we additionally preserve aggregate traffic magnitude through an additive summary,
\begin{equation}
    \mathbf{r}_{p}
    =
    \log
    \left(
    \mathbf{1}
    +
    \sum_{m\in p}
    \operatorname{softplus}
    \left(
    \operatorname{MLP}_{\mathrm{add}}
    (\mathbf{h}_m^{\mathrm M})
    \right)
    \right).
    \label{eq:phase_additive}
\end{equation}
The final phase representation combines these complementary summaries:
\begin{equation}
    \mathbf{h}_{p}^{\mathrm P}
    =
    \operatorname{MLP}_{P}
    \left(
    \operatorname{concat}
    [
    \mathbf{c}_{p},
    \mathbf{r}_{p}
    ]
    \right).
    \label{eq:phase_token}
\end{equation}
$\mathbf{h}_{p}^{\mathrm P}$ thus encodes both the movement composition of phase $p$ and its aggregate traffic magnitude.

\paragraph{Candidate Phase Scoring.}
To accommodate different numbers of feasible phases across junctions and signal plans, BEVLight applies the same scorer to every candidate. Let $p_{k_t}$ denote the current phase and define
$\Delta\mathbf{h}_{p}^{\mathrm P}
=
\mathbf{h}_{p}^{\mathrm P}
-
\mathbf{h}_{p_{k_t}}^{\mathrm P}$.
The score of candidate phase $p$ is
\begin{equation}
    s_p
    =
    \operatorname{MLP}_{D}
    \left(
    \operatorname{concat}
    [
    \mathbf{h}_{p}^{\mathrm P},
    \mathbf{h}_{p_{k_t}}^{\mathrm P},
    \Delta\mathbf{h}_{p}^{\mathrm P},
    \operatorname{Fourier}
    \left(
    \tau_t/\tau_{\max}
    \right)
    ]
    \right),
    \label{eq:phase_score}
\end{equation}
where $\tau_t$ is the elapsed duration of the current phase and $\tau_{\max}$ is the maximum phase duration.
The policy selects the highest-scoring feasible phase,
$a_t=\arg\max_{p\in\mathcal P}s_p$.
Because both phase construction and candidate scoring are shared across junctions, junction-specific signal configurations are handled through the topology rather than junction-specific output heads. 

\subsection{Human Demonstration Collection}
\label{sec:demonstration}

To supervise the shared phase scorer, we collect human traffic-control demonstrations through a web-based virtual teleoperation platform.
At each decision step, participants observe the evolving junction through a continuous BEV view, select the preferred phase from the feasible set $\mathcal P$, and rank all candidate phases.
Each demonstration therefore provides an observation $o_t$, a selected phase $a_t^*$, and a preference ranking $r_t(p)$ over $p\in\mathcal P$.
The ranking distinguishes among non-selected phases that a single action label would treat as equally undesirable.

The simulator additionally records the position of every vehicle, allowing the same traffic evolution to be re-rendered under different visual appearances while preserving the original human preferences.
These recorded states also provide traffic supervision for learning control-relevant lane representations.
Details of the annotation interface, demonstration protocol, data collection, and processing are provided in Appendix~\ref{app:human_demonstrations}.

\subsection{Policy Training}
\label{sec:training}

BEVLight is trained in two stages. The first stage learns traffic-aware lane and movement representations from simulator-provided supervision, while the second stage learns the signal-control policy from human demonstrations.

\paragraph{Stage I: Traffic-Aware Representation Pretraining.}
We attach auxiliary prediction heads to the lane and movement representations to supervise control-relevant traffic information.
At the lane level, the model predicts queue length, vehicle count, and vehicle positions along the lane; at the movement level, it predicts queue length and vehicle count.
The first-stage objective is
\begin{equation}
    \mathcal L_{\mathrm{stage\,1}}
    =
    \left(
    \mathcal L_{q}^{\mathcal L}
    +
    \mathcal L_{n}^{\mathcal L}
    +
    \mathcal L_{\mathrm{Ch}}^{\mathcal L}
    \right)
    +
    \left(
    \mathcal L_{q}^{\mathcal M}
    +
    \mathcal L_{n}^{\mathcal M}
    \right).
    \label{eq:stage1_objective}
\end{equation}
This supervision encourages the learned representations to retain traffic information relevant to downstream signal decisions. Detailed definitions of Stage I are provided in Appendix~\ref{app:training_objectives}.

\paragraph{Stage II: Human-Supervised Policy Learning.}
Starting from the pretrained representations, we train the phase scorer using both the human-selected phase and the ranking over all feasible candidates.
Let $\boldsymbol\pi_t$ denote the policy distribution obtained by softmax-normalizing the candidate scores in \Eqref{eq:phase_score}.
We construct a soft ranking target $\mathbf y_t$ from the human ranking $r_t(p)$ following the top-one listwise formulation of ListNet~\citep{cao2007learning}.
With $\mathbf e_{a_t^*}$ denoting the one-hot target for the selected phase, the second-stage objective is
\begin{equation}
    \mathcal{L}_{\mathrm{stage\,2}}
    =
    \operatorname{CE}
    (\mathbf e_{a_t^*},\boldsymbol\pi_t)
    +
    \lambda_{\mathrm{soft}}
    \operatorname{CE}
    (\mathbf y_t,\boldsymbol\pi_t).
    \label{eq:stage2_objective}
\end{equation}
The first term learns the human-selected phase, while the second preserves the relative preferences among all feasible candidates.
The construction of the ranking target and the individual training objectives are detailed in Appendix~\ref{app:training_objectives}.

\begin{figure}[t]
  \centering
  \includegraphics[width=\textwidth]{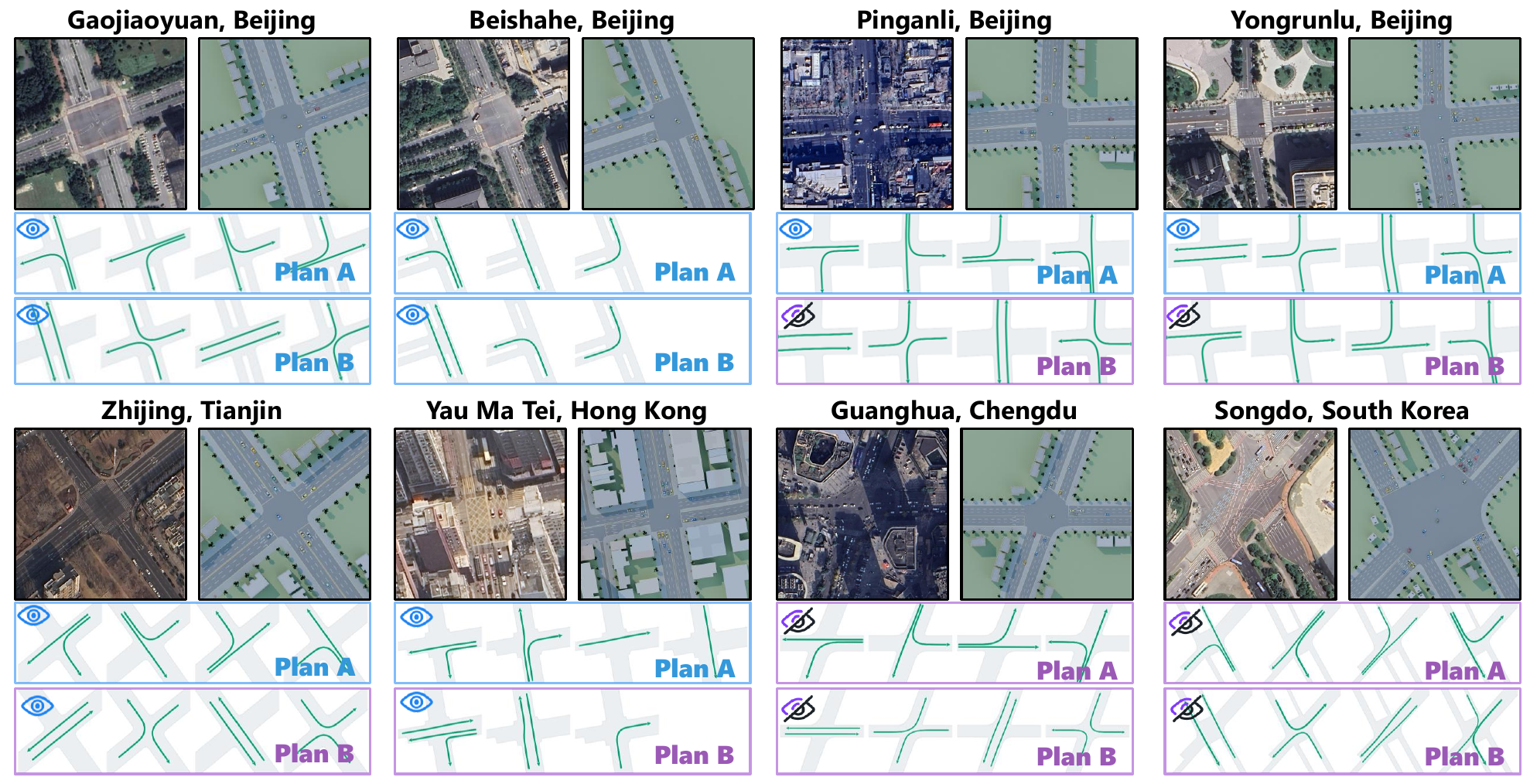}
  \caption{Eight junctions, each with two signal plans. \textcolor{trainblue}{\textbf{Blue}} borders mark plans used for training, while \textcolor{heldpurple}{\textbf{purple}} borders mark held-out plans.}
  \label{fig:junctions}
\end{figure}

\section{Experiments}
\label{sec:experiments}

\subsection{Experimental Setup}
\label{sec:experimental_setup}

\paragraph{Scenarios.}
We conduct simulation experiments in TranSimHub~\citep{wang2025transimhub}, which uses SUMO~\citep{lopez2018sumo} to simulate vehicle movements and provides an interface for rendering BEV observations (Appendix~\ref{app:experiment_settings}). \Figref{fig:junctions} shows the eight junctions used in our experiments. Each junction has two signal plans, A and B, which share the same layout but differ in their phase configurations (Appendix~\ref{app:simulated_scenes}). We consider five traffic demand patterns (Appendix~\ref{app:traffic_demand}) and evaluate three cumulative generalization settings by progressively introducing unseen demand patterns, signal plans, and junctions. For offline simulation-to-real evaluation, we use real UAV observations from the Songdo dataset \citep{fonod2025advanced}.

\paragraph{Compared Methods and Metrics.}
We compare BEVLight with two conventional controllers, FixedTime and MaxPressure~\cite{varaiya2013max}, and four structured-state RL controllers: UniTSA~\cite{wang2024unitsa}, IntelliLight~\cite{wei2018intellilight}, AttendLight~\cite{oroojlooy2020attendlight}, and MPLight~\cite{chen2020toward}. We also include two visual perception-to-control baselines. YOLO-based Count (YOLO-C) assigns detected vehicles to lanes and uses the resulting lane-level counts as inputs to AttendLight, while BEVLight-State (BEV-S) feeds BEVLight-estimated lane states into AttendLight. BEVLight directly predicts a signal phase from BEV observations and junction topology. We report average vehicle waiting time, where lower values indicate better performance. Further details on the compared methods are provided in Appendix~\ref{app:compared_methods}.

\subsection{Closed-Loop Generalization}
\label{sec:closed_loop_results}

Table~\ref{tab:closed_loop_results} reports closed-loop average vehicle waiting time under the three generalization settings. BEVLight achieves the lowest waiting time in 12 of 16 evaluation scenarios, demonstrating that human-supervised control from visual observations can match or even outperform RL-based TSC methods supplied with ground-truth traffic states. Human demonstrations provide more than a single preferred action. They rank feasible phases, expressing relative preferences among alternatives with different traffic conditions, as illustrated in Appendix~\ref{app:human_action_ranking}. This richer supervision guides the policy in comparing candidate phases rather than only reproducing the selected action.

\begin{table}[t]
  \centering
  \caption{Closed-loop average vehicle waiting time (s) across three generalization settings. Values are mean $\pm$ sample standard deviation over the two held-out demand patterns. Lower is better. $^{\ast}$, $^{\dagger}$, and $^{\ddagger}$ denote the first-, second-,
    and third-best results in each row, respectively.
  }
  \label{tab:closed_loop_results}

  \footnotesize
  \setlength{\tabcolsep}{2.2pt}
  \renewcommand{\arraystretch}{1.14}

  \resizebox{\textwidth}{!}{%
  \begin{tabular}{@{}l*{9}{c}@{}}
    \toprule
    & \multicolumn{2}{c}{Conventional}
    & \multicolumn{4}{c}{Structured-State RL}
    & \multicolumn{3}{c}{Vision-Based} \\
    \cmidrule(lr){2-3}
    \cmidrule(lr){4-7}
    \cmidrule(l){8-10}

    Junction / Plan
    & FixedTime
    & MaxPressure
    & UniTSA
    & IntelliLight
    & AttendLight
    & MPLight
    & YOLO-C
    & BEV-S
    & \textbf{BEVLight} \\
    \midrule

    \sectionheader{Unseen Demand}
    \addlinespace[0.8pt]

    Beishahe A
    & \resultvalue{18.64}{0.39}
    & \resultvalue{8.02}{1.87}
    & \resultvalue{11.67}{0.62}
    & \secondresult{7.45}{0.06}
    & \resultvalue{8.08}{0.68}
    & \resultvalue{8.26}{1.24}
    & \resultvalue{8.01}{2.07}
    & \thirdresult{7.73}{0.68}
    & \firstresult{7.40}{0.23} \\

    Beishahe B
    & \resultvalue{18.30}{1.24}
    & \resultvalue{8.01}{1.81}
    & \resultvalue{10.21}{1.38}
    & \thirdresult{6.68}{1.32}
    & \resultvalue{6.83}{1.22}
    & \secondresult{6.67}{0.93}
    & \resultvalue{7.91}{1.76}
    & \resultvalue{7.57}{2.07}
    & \firstresult{6.62}{0.60} \\

    Gaojiaoyuan A
    & \resultvalue{26.09}{3.99}
    & \resultvalue{10.20}{2.78}
    & \resultvalue{13.04}{4.15}
    & \resultvalue{10.55}{3.25}
    & \resultvalue{10.28}{4.90}
    & \resultvalue{10.71}{4.38}
    & \secondresult{9.55}{4.06}
    & \thirdresult{10.08}{4.45}
    & \firstresult{9.12}{2.31} \\

    Gaojiaoyuan B
    & \resultvalue{25.12}{3.01}
    & \resultvalue{10.37}{1.38}
    & \resultvalue{11.69}{4.13}
    & \resultvalue{9.34}{0.43}
    & \firstresult{9.18}{2.06}
    & \resultvalue{10.18}{1.81}
    & \secondresult{9.24}{0.25}
    & \resultvalue{10.07}{0.85}
    & \secondresult{9.24}{1.21} \\

    Zhijingdao A
    & \resultvalue{25.65}{0.66}
    & \resultvalue{10.21}{0.90}
    & \resultvalue{11.89}{2.05}
    & \thirdresult{9.51}{1.09}
    & \secondresult{9.42}{0.56}
    & \resultvalue{9.61}{0.71}
    & \resultvalue{9.79}{0.07}
    & \resultvalue{9.95}{1.00}
    & \firstresult{8.89}{0.50} \\

    Yau Ma Tei A
    & \resultvalue{21.00}{0.11}
    & \resultvalue{7.45}{1.03}
    & \resultvalue{8.66}{1.20}
    & \thirdresult{6.60}{0.28}
    & \secondresult{6.39}{0.58}
    & \resultvalue{6.95}{0.68}
    & \resultvalue{6.75}{1.06}
    & \resultvalue{7.04}{0.37}
    & \firstresult{6.09}{0.74} \\

    Pinganli A
    & \resultvalue{25.21}{0.05}
    & \secondresult{10.23}{0.14}
    & \resultvalue{13.80}{0.98}
    & \resultvalue{11.14}{2.33}
    & \thirdresult{10.44}{1.01}
    & \resultvalue{11.08}{1.27}
    & \resultvalue{10.52}{0.81}
    & \resultvalue{10.75}{0.80}
    & \firstresult{10.10}{0.84} \\

    Yongrunlu A
    & \resultvalue{27.27}{2.06}
    & \resultvalue{10.85}{0.06}
    & \resultvalue{12.14}{0.10}
    & \thirdresult{10.29}{0.08}
    & \secondresult{10.24}{0.88}
    & \resultvalue{10.75}{0.67}
    & \resultvalue{10.45}{0.47}
    & \resultvalue{10.41}{0.84}
    & \firstresult{9.54}{0.46} \\

    \midrule
    \sectionheader{Unseen Signal Plan + Demand}
    \addlinespace[0.8pt]

    Zhijingdao B
    & \resultvalue{26.22}{5.35}
    & \thirdresult{10.13}{1.11}
    & \resultvalue{12.62}{1.76}
    & \firstresult{9.96}{1.13}
    & \resultvalue{10.31}{0.60}
    & \resultvalue{11.00}{0.37}
    & \resultvalue{10.44}{0.28}
    & \resultvalue{10.60}{0.19}
    & \secondresult{10.09}{1.14} \\

    Yau Ma Tei B
    & \resultvalue{36.86}{1.51}
    & \resultvalue{21.21}{6.29}
    & \resultvalue{24.68}{7.49}
    & \resultvalue{21.40}{11.05}
    & \thirdresult{20.02}{7.85}
    & \secondresult{19.44}{3.71}
    & \resultvalue{20.84}{5.00}
    & \resultvalue{21.30}{9.20}
    & \firstresult{16.52}{4.67} \\

    Pinganli B
    & \resultvalue{25.62}{0.63}
    & \resultvalue{9.86}{0.78}
    & \resultvalue{11.29}{0.69}
    & \resultvalue{10.40}{1.14}
    & \firstresult{9.65}{0.64}
    & \resultvalue{9.81}{0.13}
    & \thirdresult{9.74}{0.69}
    & \resultvalue{10.24}{1.39}
    & \secondresult{9.73}{0.56} \\

    Yongrunlu B
    & \resultvalue{25.37}{3.07}
    & \resultvalue{11.14}{1.13}
    & \resultvalue{11.53}{0.08}
    & \resultvalue{10.59}{0.22}
    & \resultvalue{10.24}{0.95}
    & \resultvalue{10.51}{0.18}
    & \thirdresult{10.23}{0.23}
    & \secondresult{10.22}{0.75}
    & \firstresult{10.08}{0.15} \\

    \midrule
    \sectionheader{Unseen Junction + Plan + Demand}
    \addlinespace[0.8pt]

    Guanghua A
    & \resultvalue{22.49}{1.03}
    & \resultvalue{9.19}{1.01}
    & \resultvalue{13.39}{5.87}
    & \resultvalue{8.91}{0.04}
    & \secondresult{8.48}{0.27}
    & \resultvalue{8.85}{1.15}
    & \thirdresult{8.84}{0.98}
    & \resultvalue{9.59}{1.14}
    & \firstresult{7.95}{0.18} \\

    Guanghua B
    & \resultvalue{23.69}{0.23}
    & \resultvalue{9.12}{1.15}
    & \resultvalue{12.38}{0.51}
    & \firstresult{8.51}{0.45}
    & \secondresult{9.02}{0.94}
    & \resultvalue{9.31}{0.22}
    & \resultvalue{9.38}{0.22}
    & \resultvalue{10.04}{0.90}
    & \thirdresult{9.04}{0.62} \\

    Songdo A
    & \resultvalue{30.16}{0.58}
    & \secondresult{13.38}{0.21}
    & \resultvalue{17.23}{3.93}
    & \thirdresult{13.47}{0.13}
    & \resultvalue{13.87}{1.31}
    & \resultvalue{14.93}{1.80}
    & \resultvalue{14.46}{1.09}
    & \resultvalue{14.31}{0.75}
    & \firstresult{13.08}{1.14} \\

    Songdo B
    & \resultvalue{30.28}{0.90}
    & \resultvalue{19.47}{5.32}
    & \secondresult{18.52}{2.64}
    & \resultvalue{19.45}{7.67}
    & \resultvalue{19.16}{7.93}
    & \resultvalue{23.43}{9.87}
    & \resultvalue{19.43}{7.16}
    & \thirdresult{19.15}{4.83}
    & \firstresult{18.10}{7.13} \\

    \bottomrule
  \end{tabular}%
  }
\end{table}

\paragraph{Unseen demand.}
BEVLight achieves the lowest waiting time in seven of eight scenarios and closely matches AttendLight on Gaojiaoyuan~B (9.24~s versus 9.18~s). Despite receiving accurate simulator-provided quantities such as queue lengths, the structured-state RL methods do not consistently outperform the human-supervised visual policy. This suggests that the information available to the controller and how it is used to prioritize phases both matter for control performance. Visual observations retain vehicle positions and spatial queue patterns, while human rankings supervise how the competing phases should be prioritized. Averaged across this setting, BEVLight reduces waiting time by 7.2\% relative to YOLO-C and 9.0\% relative to BEV-S.

\paragraph{Unseen signal plans and demand.}
Across signal plans, the same action index can correspond to different movement combinations, making fixed preferences over action indices insufficient. BEVLight compares phases through the traffic served by their constituent movements, using the topology-grounded representation. On Yau Ma Tei~B, BEVLight reduces waiting time from 19.44~s for MPLight, the strongest baseline in this scenario, to 16.52~s, a 15.0\% improvement. On Zhijingdao~B and Pinganli~B, it remains within 0.13~s and 0.08~s of the best baseline, respectively.

\paragraph{Unseen junctions, signal plans, and demand.}
Finally, we evaluate BEVLight under jointly unseen junction layouts, signal plans, and demand patterns. BEVLight achieves the lowest waiting time on Guanghua~A, Songdo~A, and Songdo~B, while remaining close to AttendLight on Guanghua~B (9.04~s versus 9.02~s). It also outperforms the two-stage visual baselines in all four scenarios, with average reductions of 7.6\% over YOLO-C and 9.3\% over BEV-S. This advantage suggests that directly learning decision-relevant visual cues may alleviate the propagation of intermediate state-estimation errors into control decisions. These results demonstrate that BEVLight can transfer phase preferences learned from human rankings across changes in demand, signal plans, and junction layouts, achieving competitive closed-loop control from visual observations even against RL-based methods using ground-truth traffic states.

\subsection{Ablation Study}
\label{sec:ablation}

Table~\ref{tab:ablation} examines the contributions of adaptive pooling, visual pretraining, and soft ranking at Guanghua and Songdo. Combining all three components consistently improves control performance across these unseen settings.

\begin{wraptable}{r}{0.56\textwidth}
  \centering
  \caption{Ablation results in average waiting time (s).}
  \label{tab:ablation}
  \footnotesize
  \setlength{\tabcolsep}{3pt}
  \renewcommand{\arraystretch}{1.1}
  \begin{tabular*}{\linewidth}{@{\extracolsep{\fill}}ccc cccc@{}}
    \toprule
    \multicolumn{3}{c}{Components}
    & \multicolumn{2}{c}{Guanghua}
    & \multicolumn{2}{c}{Songdo} \\
    \cmidrule(lr){1-3}
    \cmidrule(lr){4-5}
    \cmidrule(l){6-7}
    \shortstack{Adaptive\\Pooling}& \shortstack{Visual\\Pretraining}
    & \shortstack{Soft\\Loss}
    & A & B & A & B \\
    \midrule
    \xmark & \xmark & \xmark & 9.20 & 10.73 & 13.96 & 23.79 \\
    \cmark & \xmark & \xmark & 9.38 & 10.91 & 13.08 & 29.03 \\
    \cmark & \cmark & \xmark & 8.78 & 9.12 & 12.86 & 18.20 \\
    \midrule
    \cmark & \cmark & \cmark & 7.95 & 9.04 & 12.77 & 18.10 \\
    \bottomrule
  \end{tabular*}
\end{wraptable}

Adaptive pooling alone yields mixed results, suggesting that its additional parameters are difficult to learn from action supervision without explicit perception guidance. Adding visual pretraining improves performance across all four settings, supporting the importance of perception supervision for learning useful lane-level features. Soft ranking further reduces waiting time in all scenarios. These modest gains are consistent with its role in capturing relative phase preferences when multiple actions are plausible, providing richer guidance than a single preferred action.

\begin{figure}[!t]
  \centering
  \includegraphics[width=\textwidth,pagebox=cropbox]{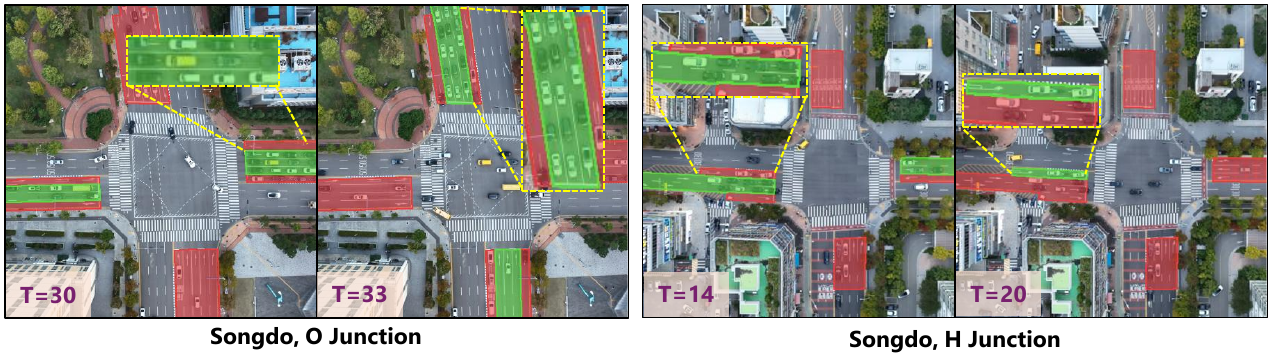}
  \caption{Phase predictions on real UAV imagery from Songdo intersections.}
  \label{fig:real_case_study}
\end{figure}

\subsection{Offline Simulation-to-Real Case Study}
\label{sec:sim_to_real}

We evaluate BEVLight on recorded UAV imagery from four real intersections in Songdo. \Figref{fig:real_case_study} illustrates BEVLight's zero-shot transfer to real-world traffic scenes through plausible phase decisions without real-world fine-tuning.

At Junction O, BEVLight selects the east-west through phase at $T=30$~s. By $T=33$~s, after east-west traffic has discharged, the model favors the north-south through phase, shifting priority to the remaining demand. Similarly, at Junction H, the model initially favors east-west through movements and then switches to the west-approach left-turn movement as through traffic subsides. These predictions qualitatively align with human control intuition, suggesting that the learned phase preferences remain meaningful under real-world visual conditions.

\subsection{Inference Time}

\begin{wraptable}{r}{0.39\textwidth}
  \centering
  \caption{Model size and full inference latency.}
  \label{tab:inference_efficiency}
  \small
  \setlength{\tabcolsep}{3pt}
  \renewcommand{\arraystretch}{1.1}
  \begin{tabular*}{\linewidth}{@{\extracolsep{\fill}}lccc@{}}
    \toprule
    Method & \shortstack{Params\\(M)} & GFLOPs
           & \shortstack{Latency\\(ms) $\downarrow$} \\
    \midrule
    YOLO11n     & 2.662 & 26.74 & 18.44 \\
    BEV-S & 1.260 & 37.61 &  8.60 \\
    BEVLight       & 1.260 & 37.61 &  8.59 \\
    \bottomrule
  \end{tabular*}
\end{wraptable}

We evaluate inference efficiency on an NVIDIA RTX 4090 with batch size 1. As shown in Table~\ref{tab:inference_efficiency}, BEVLight contains only 1.26M parameters, approximately 53\% fewer than YOLO11n. Its full inference latency is 8.59~ms, well below the 10~s control interval, indicating sufficient inference speed on the evaluated hardware.

BEVLight's compact design reflects its focus on control-relevant representations: it scores feasible phases from latent traffic features without reconstructing individual vehicle locations during inference. Although its FLOPs exceed those of YOLO11n, its latency is approximately 53\% lower, highlighting that operation counts alone do not determine runtime.

\section{Related Work}

Learning-based TSC has advanced from RL policies \citep{wei2019presslight,wei2019colight,chen2020toward,zhang2022expression,wang2024ccda} to LLM agents \citep{lai2025llmlight,yuan2026collmlight,zou2026traffic,pang2026illmtsc}. Despite their different architectures, most methods rely on structured traffic states, such as vehicle counts, queue lengths, speeds, and pressure \citep{wei2019survey}. These states are directly available in simulation but require estimation from sensor data in real deployments, introducing errors that complicate transfer. Within this setting, generalizable policies transfer across traffic demands, road topologies, and signal configurations \citep{oroojlooy2020attendlight,devailly2024model,wang2024unitsa,lai2025llmlight,zou2026traffic}, while demonstration-based methods reduce trial-and-error training using data from conventional controllers, virtual experts, or traffic engineers \citep{xiong2019demolight,wang2022demonstration,ye2023initlight,li2020deep,li2022dataaugmented}. These advances, however, remain grounded in structured states rather than learning control from images.

Vision-based TSC maps images to signal decisions without predefined traffic states. Existing studies explore end-to-end RGB control, synthetic visual data, and vision-language reasoning \citep{garg2022vision,chen2024syntrac,wang2025vlmlight}, demonstrating the potential of visual control but generally considering a fixed or limited set of junction and signal configurations. UAVs offer a flexible platform for wide-area traffic observation \citep{menouar2017uav}, yet existing studies primarily focus on traffic monitoring and trajectory extraction rather than direct signal control \citep{fonod2025advanced}. BEVLight learns a shared visual TSC policy from human demonstrations on simulated UAV observations, generalizes across heterogeneous junction geometries and signal configurations, and exhibits plausible zero-shot phase decisions on real UAV imagery.

\section{Conclusion}
\label{sec:conclusion}
We presented BEVLight, a visual traffic signal control framework that maps UAV bird's-eye-view observations directly to signal actions. BEVLight grounds visual features onto lanes and composes them along the junction topology into phase representations, so that a single shared scorer ranks candidate phases across intersections with different layouts and signal plans. The scorer is trained from human demonstrations that select and rank feasible phases, on top of representations pretrained with simulator-provided traffic states and re-rendered under diverse appearances. In closed-loop experiments, BEVLight generalizes zero-shot to unseen demands, signal plans, and junctions, matching or outperforming structured-state RL methods that receive ground-truth traffic states, and transfers to real UAV observations without target-domain adaptation. Future work includes robustness to UAV altitude and viewpoint changes and coordinated control over multiple junctions.

\bibliographystyle{unsrt}
\bibliography{reference}

@inproceedings{cao2007learning,
  title={Learning to rank: from pairwise approach to listwise approach},
  author={Cao, Zhe and Qin, Tao and Liu, Tie-Yan and Tsai, Ming-Feng and Li, Hang},
  booktitle={Proceedings of the 24th international conference on Machine learning},
  pages={129--136},
  year={2007}
}

@article{wang2025transimhub,
  title={{TranSimHub}: A Unified Air-Ground Simulation Platform for Multi-Modal Perception and Decision-Making},
  author={Wang, Maonan and Chen, Yirong and Cai, Yuxin and Pang, Aoyu and Xie, Yuejiao and Ma, Zian and Xu, Chengcheng and Jiang, Kemou and Wang, Ding and Roullet, Laurent and others},
  journal={arXiv preprint arXiv:2510.15365},
  year={2025}
}

@inproceedings{lopez2018sumo,
  title={Microscopic Traffic Simulation using {SUMO}},
  author={Lopez, Pablo Alvarez and Behrisch, Michael and Bieker-Walz, Laura and Erdmann, Jakob and Fl{"o}tter{"o}d, Yun-Pang and Hilbrich, Robert and L{"u}cken, Leonhard and Rummel, Johannes and Wagner, Peter and Wie{ss}ner, Evamarie},
  booktitle={2018 21st International Conference on Intelligent Transportation Systems (ITSC)},
  year={2018},
  organization={IEEE}
}

@article{raffin2021stable,
  title={Stable-baselines3: Reliable reinforcement learning implementations},
  author={Raffin, Antonin and Hill, Ashley and Gleave, Adam and Kanervisto, Anssi and Ernestus, Maximilian and Dormann, Noah},
  journal={Journal of machine learning research},
  volume={22},
  number={268},
  pages={1--8},
  year={2021}
}

@article{varaiya2013max,
  title={Max Pressure Control of a Network of Signalized Intersections},
  author={Varaiya, Pravin},
  journal={Transportation Research Part C: Emerging Technologies},
  volume={36},
  pages={177--195},
  year={2013},
  doi={10.1016/j.trc.2013.08.014}
}

@article{khanam2024yolov11,
  title={Yolov11: An overview of the key architectural enhancements},
  author={Khanam, Rahima and Hussain, Muhammad},
  journal={arXiv preprint arXiv:2410.17725},
  year={2024}
}

@inproceedings{wei2019presslight,
  title={Presslight: Learning max pressure control to coordinate traffic signals in arterial network},
  author={Wei, Hua and Chen, Chacha and Zheng, Guanjie and Wu, Kan and Gayah, Vikash and Xu, Kai and Li, Zhenhui},
  booktitle={Proceedings of the 25th ACM SIGKDD international conference on knowledge discovery \& data mining},
  pages={1290--1298},
  year={2019}
}

@inproceedings{chen2020toward,
  title={Toward a thousand lights: Decentralized deep reinforcement learning for large-scale traffic signal control},
  author={Chen, Chacha and Wei, Hua and Xu, Nan and Zheng, Guanjie and Yang, Ming and Xiong, Yuanhao and Xu, Kai and Li, Zhenhui},
  booktitle={Proceedings of the AAAI conference on artificial intelligence},
  volume={34},
  number={04},
  pages={3414--3421},
  year={2020}
}

@inproceedings{wei2019colight,
  title={Colight: Learning network-level cooperation for traffic signal control},
  author={Wei, Hua and Xu, Nan and Zhang, Huichu and Zheng, Guanjie and Zang, Xinshi and Chen, Chacha and Zhang, Weinan and Zhu, Yanmin and Xu, Kai and Li, Zhenhui},
  booktitle={Proceedings of the 28th ACM international conference on information and knowledge management},
  pages={1913--1922},
  year={2019}
}

@inproceedings{zhang2022expression,
  title={Expression might be enough: Representing pressure and demand for reinforcement learning based traffic signal control},
  author={Zhang, Liang and Wu, Qiang and Shen, Jun and L{\"u}, Linyuan and Du, Bo and Wu, Jianqing},
  booktitle={International Conference on Machine Learning},
  pages={26645--26654},
  year={2022},
  organization={PMLR}
}

@inproceedings{lai2025llmlight,
  title={{LLMlight}: Large language models as traffic signal control agents},
  author={Lai, Siqi and Xu, Zhao and Zhang, Weijia and Liu, Hao and Xiong, Hui},
  booktitle={Proceedings of the 31st ACM SIGKDD Conference on Knowledge Discovery and Data Mining V. 1},
  pages={2335--2346},
  year={2025}
}

@inproceedings{yuan2026collmlight,
  title={{CoLLMLight}: Cooperative large language model agents for network-wide traffic signal control},
  author={Yuan, Zirui and Lai, Siqi and Liu, Hao},
  booktitle={International Conference on Learning Representations},
  volume={2026},
  pages={66687--66715},
  year={2026}
}

@inproceedings{zou2026traffic,
  title={{Traffic-R1}: Reinforced llms bring human-like reasoning to traffic signal control systems},
  author={Zou, Xingchen and Yang, Yuhao and Chen, Zheng and Hao, Xixuan and Chen, Yiqi and Huang, Chao and Liang, Yuxuan},
  booktitle={Proceedings of the 64th Annual Meeting of the Association for Computational Linguistics (Volume 1: Long Papers)},
  pages={21823--21838},
  year={2026}
}

@article{menouar2017uav,
  title={UAV-enabled intelligent transportation systems for the smart city: Applications and challenges},
  author={Menouar, Hamid and Guvenc, Ismail and Akkaya, Kemal and Uluagac, A Selcuk and Kadri, Abdullah and Tuncer, Adem},
  journal={IEEE Communications Magazine},
  volume={55},
  number={3},
  pages={22--28},
  year={2017},
  publisher={IEEE}
}

@article{devailly2024model,
  title={Model-based graph reinforcement learning for inductive traffic signal control},
  author={Devailly, Fran{\c{c}}ois-Xavier and Larocque, Denis and Charlin, Laurent},
  journal={IEEE Open Journal of Intelligent Transportation Systems},
  volume={5},
  pages={238--250},
  year={2024},
  publisher={IEEE}
}

@ARTICLE{wang2024unitsa,
  author={Wang, Maonan and Xiong, Xi and Kan, Yuheng and Xu, Chengcheng and Pun, Man-On},
  journal={IEEE Transactions on Vehicular Technology}, 
  title={{UniTSA}: A Universal Reinforcement Learning Framework for V2X Traffic Signal Control}, 
  year={2024},
  volume={73},
  number={10},
  pages={14354-14369},
  doi={10.1109/TVT.2024.3403879}
}

@article{oroojlooy2020attendlight,
  title={Attendlight: Universal attention-based reinforcement learning model for traffic signal control},
  author={Oroojlooy, Afshin and Nazari, Mohammadreza and Hajinezhad, Davood and Silva, Jorge},
  journal={Advances in Neural Information Processing Systems},
  volume={33},
  pages={4079--4090},
  year={2020}
}

@inproceedings{wei2018intellilight,
  title={Intellilight: A reinforcement learning approach for intelligent traffic light control},
  author={Wei, Hua and Zheng, Guanjie and Yao, Huaxiu and Li, Zhenhui},
  booktitle={Proceedings of the 24th ACM SIGKDD international conference on knowledge discovery \& data mining},
  pages={2496--2505},
  year={2018}
}

@inproceedings{wang2025vlmlight,
 author = {Wang, Maonan and Chen, Yirong and Pang, Aoyu and Cai, Yuxin and Chen, Chung Shue and Kan, Yuheng and Pun, Man On},
 booktitle = {Advances in Neural Information Processing Systems},
 editor = {D. Belgrave and C. Zhang and H. Lin and R. Pascanu and P. Koniusz and M. Ghassemi and N. Chen},
 pages = {39590--39621},
 publisher = {Curran Associates, Inc.},
 title = {{VLMLight}: Safety-Critical Traffic Signal Control via Vision-Language Meta-Control and Dual-Branch Reasoning Architecture},
 volume = {38},
 year = {2025}
}

@inproceedings{chen2024syntrac,
  title={{SynTrac}: a synthetic dataset for traffic signal control from traffic monitoring cameras},
  author={Chen, Tiejin and Shirke, Prithvi and Chakravarthi, Bharatesh and Vaghela, Arpitsinh and Da, Longchao and Lu, Duo and Yang, Yezhou and Wei, Hua},
  booktitle={2024 IEEE 27th International Conference on Intelligent Transportation Systems (ITSC)},
  pages={2386--2391},
  year={2024},
  organization={IEEE}
}

@article{fonod2025advanced,
  title={Advanced computer vision for extracting georeferenced vehicle trajectories from drone imagery},
  author={Fonod, Robert and Cho, Haechan and Yeo, Hwasoo and Geroliminis, Nikolas},
  journal={Transportation Research Part C: Emerging Technologies},
  volume={178},
  pages={105205},
  year={2025},
  publisher={Elsevier}
}

@ARTICLE{pang2026illmtsc,
  author={Pang, Aoyu and Wang, Maonan and Pun, Man-On and Chen, Chung Shue and Xiong, Xi},
  journal={IEEE Transactions on Vehicular Technology}, 
  title={{iLLM-TSC}: Integration Reinforcement Learning and Large Language Model for Traffic Signal Control Policy Improvement}, 
  year={2026},
  volume={75},
  number={8},
  pages={15762-15776},
  doi={10.1109/TVT.2026.3674284}
}

@article{li2020deep,
  title={Deep behavioral cloning for traffic control with virtual expert demonstration under a parallel learning framework},
  author={Li, Xiaoshuang and Zhu, Fenghua and Wang, Fei-Yue},
  journal={IFAC-PapersOnLine},
  volume={53},
  number={5},
  pages={176--181},
  year={2020},
  publisher={Elsevier}
}

@article{wei2019survey,
  title={A survey on traffic signal control methods},
  author={Wei, Hua and Zheng, Guanjie and Gayah, Vikash and Li, Zhenhui},
  journal={arXiv preprint arXiv:1904.08117},
  year={2019}
}

@inproceedings{garg2022vision,
  author = {Garg, Deepeka and Chli, Maria and Vogiatzis, George},
  title = {Fully-Autonomous, Vision-based Traffic Signal Control: From Simulation to Reality},
  year = {2022},
  isbn = {9781450392136},
  publisher = {International Foundation for Autonomous Agents and Multiagent Systems},
  address = {Richland, SC},
  booktitle = {Proceedings of the 21st International Conference on Autonomous Agents and Multiagent Systems},
  pages = {454–462},
  numpages = {9},
  location = {Virtual Event, New Zealand},
  series = {AAMAS '22}
}

@inproceedings{xiong2019demolight,
  title={Learning traffic signal control from demonstrations},
  author={Xiong, Yuanhao and Zheng, Guanjie and Xu, Kai and Li, Zhenhui},
  booktitle={Proceedings of the 28th ACM international conference on information and knowledge management},
  pages={2289--2292},
  year={2019}
}

@inproceedings{wang2022demonstration,
  title={Urban Traffic Signal Control with Reinforcement Learning from Demonstration Data},
  author={Wang, Min and Wu, Libing and Li, Jianxin and Wu, Dan and Ma, Chao},
  booktitle={2022 International Joint Conference on Neural Networks (IJCNN)},
  pages={1--8},
  year={2022},
  doi={10.1109/IJCNN55064.2022.9892538}
}

@inproceedings{ye2023initlight,
  title={InitLight: Initial Model Generation for Traffic Signal Control Using Adversarial Inverse Reinforcement Learning},
  author={Ye, Yutong and Zhou, Yingbo and Ding, Jiepin and Wang, Ting and Chen, Mingsong and Lian, Xiang},
  booktitle={Proceedings of the Thirty-Second International Joint Conference on Artificial Intelligence},
  pages={4949--4958},
  year={2023},
  doi={10.24963/ijcai.2023/550}
}

@article{li2022dataaugmented,
  title={Data Augmented Deep Behavioral Cloning for Urban Traffic Control Operations Under a Parallel Learning Framework},
  author={Li, Xiaoshuang and Ye, Peijun and Jin, Junchen and Zhu, Fenghua and Wang, Fei-Yue},
  journal={IEEE Transactions on Intelligent Transportation Systems},
  volume={23},
  number={6},
  pages={5128--5137},
  year={2022},
  doi={10.1109/TITS.2020.3048151}
}

@ARTICLE{wang2024ccda,
  author={Wang, Maonan and Chen, Yirong and Kan, Yuheng and Xu, Chengcheng and Lepech, Michael and Pun, Man-On and Xiong, Xi},
  journal={IEEE Transactions on Intelligent Transportation Systems}, 
  title={Traffic Signal Cycle Control With Centralized Critic and Decentralized Actors Under Varying Intervention Frequencies}, 
  year={2024},
  volume={25},
  number={12},
  pages={20085-20104},
  doi={10.1109/TITS.2024.3462153}
}

\appendix
\clearpage
In this appendix, we provide supplementary material for BEVLight, covering implementation and training details, experimental settings, human demonstration collection, additional experimental results, and broader discussion. The appendix is organized as follows:
\begin{itemize}[leftmargin=*]
    \item Appendix~\ref{app:method_details} provides implementation and training details of BEVLight.
    \item Appendix~\ref{app:experimental_setup} describes the experimental setup and comparison methods.
    \item Appendix~\ref{app:human_demonstrations} describes the collection and processing of human traffic-control demonstrations.
    \item Appendix~\ref{app:real_actions} presents additional real-world case studies.
    \item Appendix~\ref{app:discussion} discusses practical considerations, broader impact, and limitations of BEVLight.
\end{itemize}

\section{BEVLight Model Architecture and Algorithm}
\label{app:method_details}

This section provides additional details of the BEVLight architecture and training objectives introduced in Sec.~\ref{sec:method}. We first describe the complete computation flow of the topology-grounded BEV-to-phase representation, and then elaborate on the traffic-aware and human-supervised objectives used in the two-stage training procedure.

\subsection{BEVLight Architecture}
\label{app:architecture}

This subsection provides the complete architectural formulation of the topology-grounded BEV-to-phase representation introduced in Sec.~\ref{sec:representation}. Starting from a sequence of BEV observations, BEVLight first grounds visual patch features onto individual lanes, models their temporal evolution and interactions across lanes, and then follows the junction topology to progressively construct movement and phase representations. The resulting candidate phases are finally evaluated by a shared scoring function.

\paragraph{BEV-to-Lane Grounding.}
For each frame $I_j^{\mathrm{BEV}}$ in the observation sequence $\mathbf{I}_t^{\mathrm{BEV}}$, the visual encoder produces $U$ patch features,
\begin{equation}
    \mathbf{F}_j
    =
    E_{\mathrm{vis}}(I_j^{\mathrm{BEV}})
    =
    [\mathbf{f}_1^j,\ldots,\mathbf{f}_U^j],
    \label{eq:app_visual_features}
\end{equation}
where $\mathbf{f}_u^j$ denotes the feature of patch $u$.

For each lane $\ell$, its mask $\mathbf{M}_\ell$ defines the geometric correspondence between the lane and the visual patches. We characterize each lane--patch pair using two quantities. The coverage coefficient $c_{\ell,u}\in[0,1]$ measures the fraction of patch $u$ covered by $\mathbf{M}_\ell$, while the normalized longitudinal offset
\begin{equation}
    o_{\ell,u}
    =
    \frac{d_{\ell,u}}{L_\ell}
    \label{eq:app_lane_offset}
\end{equation}
specifies its position along the lane, where $d_{\ell,u}$ is the patch-to-stop-line offset and $L_\ell$ is the lane length.

Rather than assigning fixed weights according to the lane mask, BEVLight predicts a content-dependent correction for each lane--patch pair. Specifically, a two-layer MLP takes the patch feature together with its geometric attributes and combines the resulting correction with a mask-derived geometric prior:
\begin{equation}
    a_{\ell,u}^{j}
    =
    \underbrace{
    \log(c_{\ell,u}+\epsilon)
    }_{\text{geometric mask prior}}
    +
    \underbrace{
    \operatorname{MLP}_{\mathrm{pool}}
    \left(
    \operatorname{concat}
    [\mathbf{f}_u^j,c_{\ell,u},o_{\ell,u}]
    \right)
    }_{\text{content-dependent correction}},
    \label{eq:app_lane_pool_score}
\end{equation}
where $\epsilon>0$ ensures numerical stability. The scores are normalized only over patches that overlap with lane $\ell$,
\begin{equation}
    \alpha_{\ell,u}^{j}
    =
    \operatorname{Softmax}_{u:c_{\ell,u}>0}
    \left(
    a_{\ell,u}^{j}
    \right).
    \label{eq:app_lane_pool_weight}
\end{equation}

We then aggregate the position-aware patch features into a frame-level lane token,
\begin{equation}
    \mathbf{v}_{\ell}^{j}
    =
    \sum_{u}
    \alpha_{\ell,u}^{j}
    \left(
    \mathbf{f}_u^j
    +
    \operatorname{Fourier}(o_{\ell,u})
    \right).
    \label{eq:app_lane_token}
\end{equation}
The longitudinal offset therefore influences the lane representation in two complementary ways: it conditions the adaptive pooling weights in Eq.~\ref{eq:app_lane_pool_score}, and its Fourier encoding preserves within-lane positional information in the aggregated feature.

\paragraph{Temporal and Inter-Lane Modeling.}
Applying the grounding operation to all $H$ frames gives a token history
$[\mathbf{v}_{\ell}^{t-H+1},\ldots,\mathbf{v}_{\ell}^{t}]$
for each lane $\ell$. To explicitly expose short-term traffic changes, we compute the difference between adjacent lane tokens,
\begin{equation}
    \Delta\mathbf{v}_{\ell}^{j}
    =
    \mathbf{v}_{\ell}^{j}
    -
    \mathbf{v}_{\ell}^{j-1},
    \qquad
    j=t-H+2,\ldots,t,
    \label{eq:app_temporal_difference}
\end{equation}
and set
$\Delta\mathbf{v}_{\ell}^{t-H+1}=\mathbf{0}$.

A temporal Transformer independently summarizes the history of each lane:
\begin{equation}
    \mathbf{h}_{\ell}^{\mathrm L}
    =
    \operatorname{Transformer}_{\mathrm{temp}}
    \left(
    [
    \operatorname{concat}
    [
    \mathbf{v}_{\ell}^{j},
    \Delta\mathbf{v}_{\ell}^{j}
    ]
    ]_{j=t-H+1}^{t}
    \right).
    \label{eq:app_temporal_lane}
\end{equation}
Collecting the temporal representations of all lanes gives
\begin{equation}
    \mathbf{H}_{\mathcal L}
    =
    [\mathbf{h}_{\ell}^{\mathrm L}]_{\ell\in\mathcal L}.
    \label{eq:app_lane_set}
\end{equation}

We subsequently apply multi-head self-attention across lanes,
\begin{equation}
    \widetilde{\mathbf{H}}_{\mathcal L}
    =
    \operatorname{SelfAttn}_{L}
    \left(
    \mathbf{H}_{\mathcal L}
    \right),
    \label{eq:app_lane_attention}
\end{equation}
allowing each lane to incorporate information from other traffic streams at the junction. The row
$\widetilde{\mathbf h}_{\ell}^{\mathrm L}$
of
$\widetilde{\mathbf H}_{\mathcal L}$
denotes the resulting contextualized representation of lane $\ell$.

\paragraph{Lane-to-Movement Composition.}
We next propagate the contextualized lane representations through the junction topology. The incoming and outgoing lane-to-movement relations are encoded by two fixed binary incidence matrices,
$\mathbf{B}_{\mathcal L\rightarrow\mathcal M}^{\mathrm{in}}$
and
$\mathbf{B}_{\mathcal L\rightarrow\mathcal M}^{\mathrm{out}}$,
respectively. Each entry indicates whether a lane belongs to the incoming or outgoing side of a movement.

The corresponding movement-level summaries are obtained through
\begin{equation}
    \mathbf{H}_{\mathcal M}^{\mathrm{in}}
    =
    \mathbf{B}_{\mathcal L\rightarrow\mathcal M}^{\mathrm{in}}
    \widetilde{\mathbf{H}}_{\mathcal L},
    \qquad
    \mathbf{H}_{\mathcal M}^{\mathrm{out}}
    =
    \mathbf{B}_{\mathcal L\rightarrow\mathcal M}^{\mathrm{out}}
    \widetilde{\mathbf{H}}_{\mathcal L}.
    \label{eq:app_movement_aggregation}
\end{equation}
For movement $m$, the rows
$\mathbf{h}_{m}^{\mathrm{M,in}}$
and
$\mathbf{h}_{m}^{\mathrm{M,out}}$
summarize its incoming and outgoing lane context.

A shared MLP combines these two summaries with their difference:
\begin{equation}
    \bar{\mathbf{h}}_{m}^{\mathrm M}
    =
    \operatorname{MLP}_{M}
    \left(
    \operatorname{concat}
    [
    \mathbf{h}_{m}^{\mathrm{M,in}},
    \mathbf{h}_{m}^{\mathrm{M,out}},
    \mathbf{h}_{m}^{\mathrm{M,in}}
    -
    \mathbf{h}_{m}^{\mathrm{M,out}}
    ]
    \right).
    \label{eq:app_movement_token}
\end{equation}
The difference term provides a pressure-like cue that reflects the imbalance between upstream and downstream traffic conditions associated with the movement.

We further allow different movements to exchange information through self-attention:
\begin{equation}
    \mathbf{H}_{\mathcal M}
    =
    \operatorname{SelfAttn}_{M}
    \left(
    \bar{\mathbf{H}}_{\mathcal M}
    \right),
    \qquad
    \bar{\mathbf{H}}_{\mathcal M}
    =
    [\bar{\mathbf{h}}_{m}^{\mathrm M}]_{m\in\mathcal M}.
    \label{eq:app_movement_attention}
\end{equation}
The row $\mathbf{h}_{m}^{\mathrm M}$ of $\mathbf{H}_{\mathcal M}$ is the resulting contextualized representation of movement $m$.

\paragraph{Movement-to-Phase Composition.}
The movement-to-phase membership is encoded by the fixed binary incidence matrix
$\mathbf{B}_{\mathcal M\rightarrow\mathcal P}$.
For each feasible phase $p\in\mathcal P$, let
\begin{equation}
    \mathbf{H}_{p}
    =
    [\mathbf{h}_{m}^{\mathrm M}]_{m\in p}
    \label{eq:app_phase_movements}
\end{equation}
collect the representations of the movements served by phase $p$.

A shared learnable phase query $\mathbf{q}_{P}$ attends to these constituent movements:
\begin{equation}
    \mathbf{c}_{p}
    =
    \operatorname{CrossAttn}
    \left(
    \mathbf{q}_{P},
    \mathbf{H}_{p},
    \mathbf{H}_{p}
    \right).
    \label{eq:app_phase_attention}
\end{equation}
The attention summary $\mathbf{c}_{p}$ captures the relative composition of the movements within phase $p$. However, since the attention weights are normalized to sum to one, it does not directly preserve their aggregate traffic magnitude.

To retain this complementary information, each movement representation is projected to a non-negative vector,
\begin{equation}
    \mathbf{r}_{m}
    =
    \operatorname{softplus}
    \left(
    \operatorname{MLP}_{\mathrm{add}}
    (\mathbf{h}_{m}^{\mathrm M})
    \right),
    \label{eq:app_movement_additive}
\end{equation}
which is accumulated over the movements belonging to phase $p$:
\begin{equation}
    \mathbf{r}_{p}
    =
    \log
    \left(
    \mathbf{1}
    +
    \sum_{m\in p}
    \mathbf{r}_{m}
    \right).
    \label{eq:app_phase_additive}
\end{equation}
The element-wise logarithm controls the scale of the additive summary.

Finally, the attention-based composition summary and additive traffic summary are combined to form the phase representation:
\begin{equation}
    \mathbf{h}_{p}^{\mathrm P}
    =
    \operatorname{MLP}_{P}
    \left(
    \operatorname{concat}
    [
    \mathbf{c}_{p},
    \mathbf{r}_{p}
    ]
    \right).
    \label{eq:app_phase_token}
\end{equation}
The resulting
$\mathbf{h}_{p}^{\mathrm P}$
therefore captures both the relative composition of the movements served by phase $p$ and their aggregate traffic state.

\paragraph{Candidate Phase Scoring.}
BEVLight applies the same scoring function to every feasible candidate phase, allowing the policy to operate with signal plans containing different numbers of phases.

Let $p_{k_t}$ denote the currently active phase. For candidate phase $p$, we compute
\begin{equation}
    \Delta\mathbf{h}_{p}^{\mathrm P}
    =
    \mathbf{h}_{p}^{\mathrm P}
    -
    \mathbf{h}_{p_{k_t}}^{\mathrm P},
    \label{eq:app_phase_difference}
\end{equation}
which describes the representation change associated with switching from the current phase to the candidate.

The shared scorer jointly considers the candidate representation, the current phase representation, their difference, and the elapsed duration of the current phase:
\begin{equation}
    s_p
    =
    \operatorname{MLP}_{D}
    \left(
    \operatorname{concat}
    [
    \mathbf{h}_{p}^{\mathrm P},
    \mathbf{h}_{p_{k_t}}^{\mathrm P},
    \Delta\mathbf{h}_{p}^{\mathrm P},
    \operatorname{Fourier}
    \left(
    \tau_t/\tau_{\max}
    \right)
    ]
    \right),
    \label{eq:app_phase_score}
\end{equation}
where $\tau_t$ denotes the elapsed duration of the current phase and $\tau_{\max}$ is the maximum phase duration.

At each decision step, BEVLight selects
\begin{equation}
    a_t
    =
    \arg\max_{p\in\mathcal P}
    s_p.
    \label{eq:app_action_selection}
\end{equation}
Because the lane-to-movement and movement-to-phase relations are supplied through junction-specific topology while all learnable transformations are shared, the same architecture can process junctions with different lane layouts, movement sets, and feasible signal phases.

\subsection{Training Objectives and Details}
\label{app:training_objectives}

This subsection provides additional details of the two-stage training procedure introduced in Sec.~\ref{sec:training}. Stage I uses simulator-provided traffic states to learn traffic-aware lane and movement representations, while Stage II trains the phase-selection policy from human traffic-control demonstrations.

\paragraph{Stage I: Traffic-Aware Representation Pretraining.}
We apply auxiliary prediction heads to the contextualized lane representations
$\widetilde{\mathbf h}_{\ell}^{\mathrm L}$
and movement representations
$\mathbf h_m^{\mathrm M}$.
These objectives encourage the intermediate representations to preserve traffic information directly relevant to downstream signal-control decisions.

At the lane level, BEVLight predicts queue length and vehicle count. Let
$q_{\ell}$ and $n_{\ell}$ denote their ground-truth values, and
$\hat q_{\ell}$ and $\hat n_{\ell}$ the corresponding predictions. We use L2 regression objectives,
\begin{equation}
    \mathcal L_{q}^{\mathcal L}
    =
    \frac{1}{|\mathcal L|}
    \sum_{\ell\in\mathcal L}
    \left\|
    \hat q_{\ell}
    -
    q_{\ell}
    \right\|_2^2,
    \qquad
    \mathcal L_{n}^{\mathcal L}
    =
    \frac{1}{|\mathcal L|}
    \sum_{\ell\in\mathcal L}
    \left\|
    \hat n_{\ell}
    -
    n_{\ell}
    \right\|_2^2.
    \label{eq:app_lane_state_loss}
\end{equation}

The lane head additionally predicts slotwise vehicle presence and longitudinal position. Summing the predicted presence values yields the corresponding vehicle-count estimate. To preserve fine-grained spatial information, we supervise the predicted and observed vehicle positions using a bidirectional Chamfer objective
$\mathcal L_{\mathrm{Ch}}^{\mathcal L}$.
Each observed vehicle position is matched to its nearest predicted position and vice versa, with the contribution of each predicted position weighted by its predicted presence. This objective encourages
$\widetilde{\mathbf h}_{\ell}^{\mathrm L}$
to retain the locations of individual vehicles relative to the stop line rather than only aggregate traffic statistics.

At the movement level, BEVLight predicts queue length and vehicle count from
$\mathbf h_m^{\mathrm M}$.
Let $q_m$ and $n_m$ denote their ground-truth values, with predictions
$\hat q_m$ and $\hat n_m$. The corresponding objectives are
\begin{equation}
    \mathcal L_{q}^{\mathcal M}
    =
    \frac{1}{|\mathcal M|}
    \sum_{m\in\mathcal M}
    \left\|
    \hat q_m
    -
    q_m
    \right\|_2^2,
    \qquad
    \mathcal L_{n}^{\mathcal M}
    =
    \frac{1}{|\mathcal M|}
    \sum_{m\in\mathcal M}
    \left\|
    \hat n_m
    -
    n_m
    \right\|_2^2.
    \label{eq:app_movement_state_loss}
\end{equation}

The complete first-stage objective is
\begin{equation}
    \mathcal L_{\mathrm{stage\,1}}
    =
    \left(
    \mathcal L_{q}^{\mathcal L}
    +
    \mathcal L_{n}^{\mathcal L}
    +
    \mathcal L_{\mathrm{Ch}}^{\mathcal L}
    \right)
    +
    \left(
    \mathcal L_{q}^{\mathcal M}
    +
    \mathcal L_{n}^{\mathcal M}
    \right).
    \label{eq:app_stage1_objective}
\end{equation}

The simulator records the position of every vehicle at each time step. These recorded states provide the supervision above and also allow the same traffic evolution to be rendered under different visual appearances while retaining identical traffic-state targets.

\paragraph{Stage II: Human-Supervised Policy Learning.}
Starting from the representations learned in Stage I, we train the phase-selection policy using the human demonstrations described in Sec.~\ref{sec:demonstration}. Each demonstration provides the preferred phase $a_t^*$ together with a ranking $r_t(p)$ over every feasible phase $p\in\mathcal P$, where rank one denotes the most preferred phase.

Given the candidate scores $s_p$ from Eq.~\ref{eq:phase_score}, we first obtain the policy distribution
\begin{equation}
    \pi_t(p)
    =
    \frac{
    \exp(s_p)
    }{
    \sum_{p'\in\mathcal P}
    \exp(s_{p'})
    }.
    \label{eq:app_policy_distribution}
\end{equation}

The selected phase provides a hard supervision target
$\mathbf e_{a_t^*}$.
To additionally preserve the preferences among the remaining candidates, we construct a soft target from the complete human ranking. Following the top-one listwise formulation of ListNet~\citep{cao2007learning}, we define
\begin{equation}
    y_t(p)
    =
    \frac{
    \exp\left(
    -r_t(p)/|\mathcal P|
    \right)
    }{
    \sum_{p'\in\mathcal P}
    \exp\left(
    -r_t(p')/|\mathcal P|
    \right)
    }.
    \label{eq:app_ranking_target}
\end{equation}
Scaling the ranks by the number of feasible phases keeps the range of the target logits comparable across signal plans containing different numbers of candidates.

The second-stage objective jointly supervises the selected phase and the complete phase ranking:
\begin{equation}
    \mathcal L_{\mathrm{stage\,2}}
    =
    \operatorname{CE}
    \left(
    \mathbf e_{a_t^*},
    \boldsymbol\pi_t
    \right)
    +
    \lambda_{\mathrm{soft}}
    \operatorname{CE}
    \left(
    \mathbf y_t,
    \boldsymbol\pi_t
    \right).
    \label{eq:app_stage2_objective}
\end{equation}
The first term directly learns the human-selected phase, while the second transfers the relative preferences among all feasible candidates.

Because the ranking supervision is defined over the feasible phases of each individual signal plan, the same objective can be applied across junctions with different numbers and compositions of signal phases. Moreover, re-rendered observations corresponding to the same recorded traffic evolution retain the original human ranking because the underlying traffic state is unchanged.

\section{Experimental Setup}
\label{app:experimental_setup}

\subsection{Experimental Settings}
\label{app:experiment_settings}

We conduct closed-loop experiments in TranSimHub~\citep{wang2025transimhub}, with microscopic vehicle dynamics simulated by SUMO~\citep{lopez2018sumo}. All simulated roads have a speed limit of 13.9\,m/s (approximately 50\,km/h). Vehicle speeds follow a Gaussian distribution with a mean of 10\,m/s and a standard deviation of 3\,m/s, representing typical urban traffic conditions.

For the visual encoder, we employ a lightweight ResNet trained from scratch. It consists of a stride-4 convolutional stem followed by two residual blocks and produces 128-channel feature maps at 1/4 input resolution. Each observation contains five consecutive 448×448 RGB BEV frames.

All experiments are conducted on a workstation running Ubuntu~22.04 with an Intel Xeon Gold 6133 CPU, 128\,GB of memory, and an NVIDIA GeForce RTX~4090 GPU. Structured-state RL baselines are implemented with Stable-Baselines3~\citep{raffin2021stable} and trained for $10^6$ environment steps. UniTSA~\citep{wang2024unitsa} is trained with PPO, while IntelliLight~\citep{wei2018intellilight}, AttendLight~\citep{oroojlooy2020attendlight}, and MPLight~\citep{chen2020toward} are trained with DQN. Their optimization settings are summarized below.

\paragraph{PPO.}
For UniTSA~\citep{wang2024unitsa}, we use 16 parallel environments with 128 steps per rollout, resulting in 2,048 transitions per update. Each rollout is optimized for 10 epochs with a minibatch size of 64. The learning rate is $3\times10^{-4}$, with a discount factor of $0.99$ and a GAE parameter of $0.95$. We use a policy clipping range of $0.2$, disable value-function clipping, and enable advantage normalization. The entropy and value-loss coefficients are set to $0$ and $0.5$, respectively, with a maximum gradient norm of $0.5$.

\paragraph{DQN.}
For IntelliLight, AttendLight, and MPLight~\citep{wei2018intellilight, oroojlooy2020attendlight, chen2020toward}, we use a learning rate of $10^{-4}$ and a discount factor of $0.99$. The replay buffer stores up to $10^6$ transitions, with learning starting after 100 transitions. Each update performs 16 gradient steps with a batch size of 32. We use one-step returns and hard target-network updates ($\tau=1$) every 10,000 transitions. The exploration rate is linearly annealed from $1.0$ to $0.05$ over the first 10\% of training. Gradients are clipped to a maximum norm of $10$.

\subsection{Simulation Scenarios}
\label{app:simulated_scenes}

\begin{figure}[t]
  \centering
  \includegraphics[width=\textwidth]{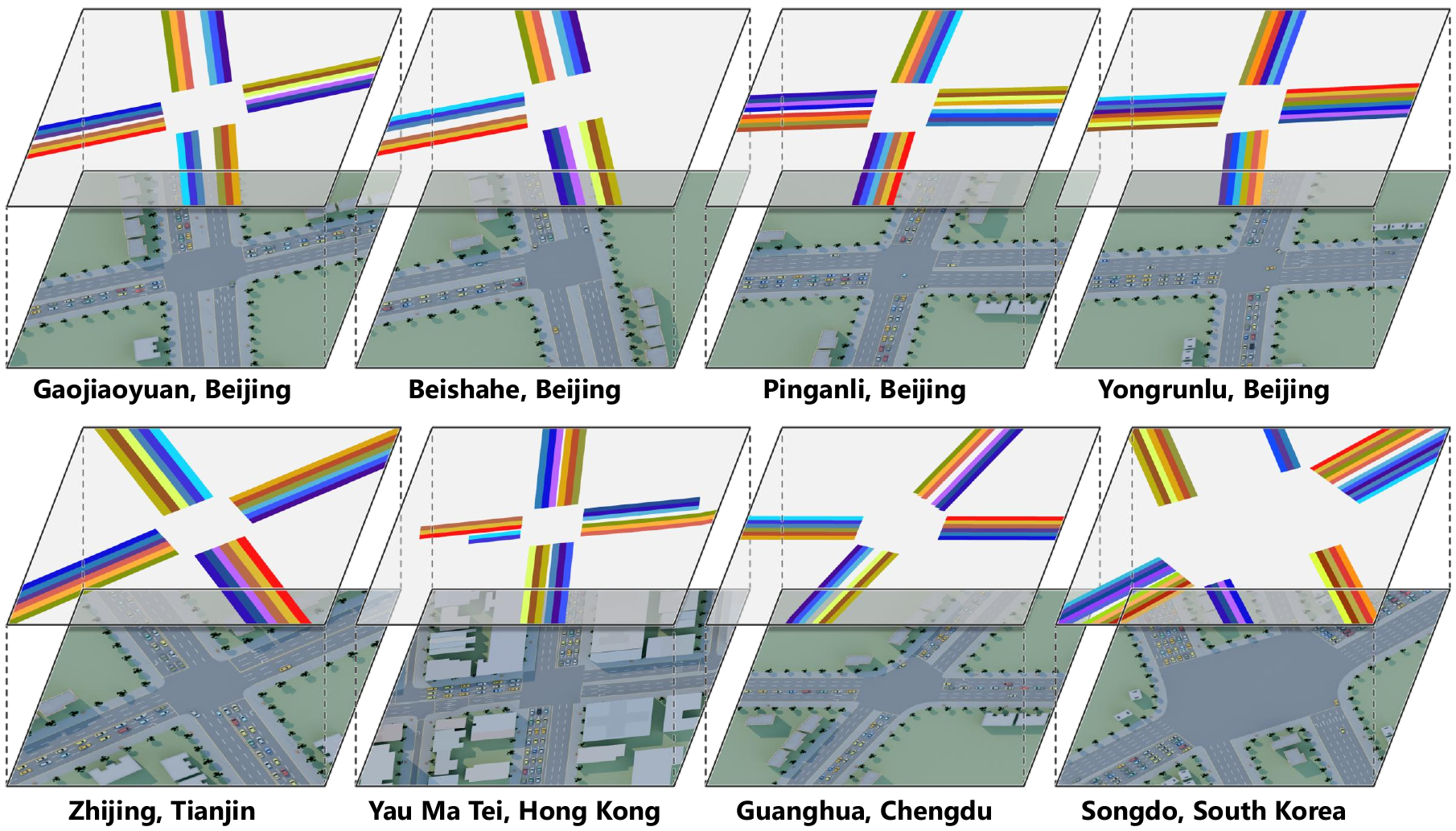}
  \caption{BEV observations and lane masks for the eight simulated intersections. For each scene, the lane masks indicate the visible lane regions used by BEVLight, with different colors denoting individual lanes.}
  \label{fig:appendix_sim_lane_mask}
\end{figure}

\begin{table}[t]
    \centering
    \caption{Spatial coverage of the eight simulated intersections.}
    \label{tab:simulation_spatial_coverage}
    \small
    \setlength{\tabcolsep}{4.5pt}
    \begin{tabular}{@{}lccccc@{}}
        \toprule
        & \multicolumn{2}{c}{Ground coverage}
        & \multicolumn{3}{c}{Visible lane length (m)} \\
        \cmidrule(lr){2-3}
        \cmidrule(lr){4-6}
        Intersection & Footprint (m $\times$ m) & Area (m$^2$) & Min. & Mean & Max. \\
        \midrule
        Gaojiaoyuan & $140.86 \times 140.86$ & 19,842 & 58.13 & 60.00 & 62.09 \\
        Beishahe & $131.48 \times 131.48$ & 17,287 & 50.82 & 57.90 & 62.87 \\
        Pinganli & $147.89 \times 147.89$ & 21,871 & 55.22 & 58.27 & 60.67 \\
        Yongrunlu & $147.50 \times 147.50$ & 21,756 & 56.59 & 59.09 & 61.20 \\
        Zhijingdao & $111.95 \times 111.95$ & 12,533 & 56.01 & 60.13 & 64.10 \\
        Yau Ma Tei & $148.48 \times 148.48$ & 22,046 & 46.29 & 56.89 & 61.69 \\
        Guanghua & $157.07 \times 157.07$ & 24,671 & 57.74 & 60.41 & 64.05 \\
        Songdo& $155.51 \times 155.51$ & 24,183 & 42.76 & 58.07 & 69.98 \\
        \bottomrule
    \end{tabular}
\end{table}

We construct eight simulation scenes from real-world intersection layouts across Beijing, Tianjin, Hong Kong, Chengdu, and Songdo, covering diverse three- and four-arm junction geometries and lane configurations. The scenes represent different urban contexts, ranging from dense inner-city intersections, such as Yau Ma Tei in Hong Kong, to newly developed districts with wider road layouts, such as Songdo.

\Figref{fig:appendix_sim_lane_mask} visualizes the rendered BEV observation and corresponding lane masks for each intersection. The masks define the lane regions used by BEVLight and illustrate the visible extent of both incoming and outgoing lanes under the UAV viewpoint. Table~\ref{tab:simulation_spatial_coverage} further reports the ground coverage and median visible lane length at a simulated UAV height of 100\,m.

\begin{figure*}[t]
    \centering
    \includegraphics[width=\textwidth]{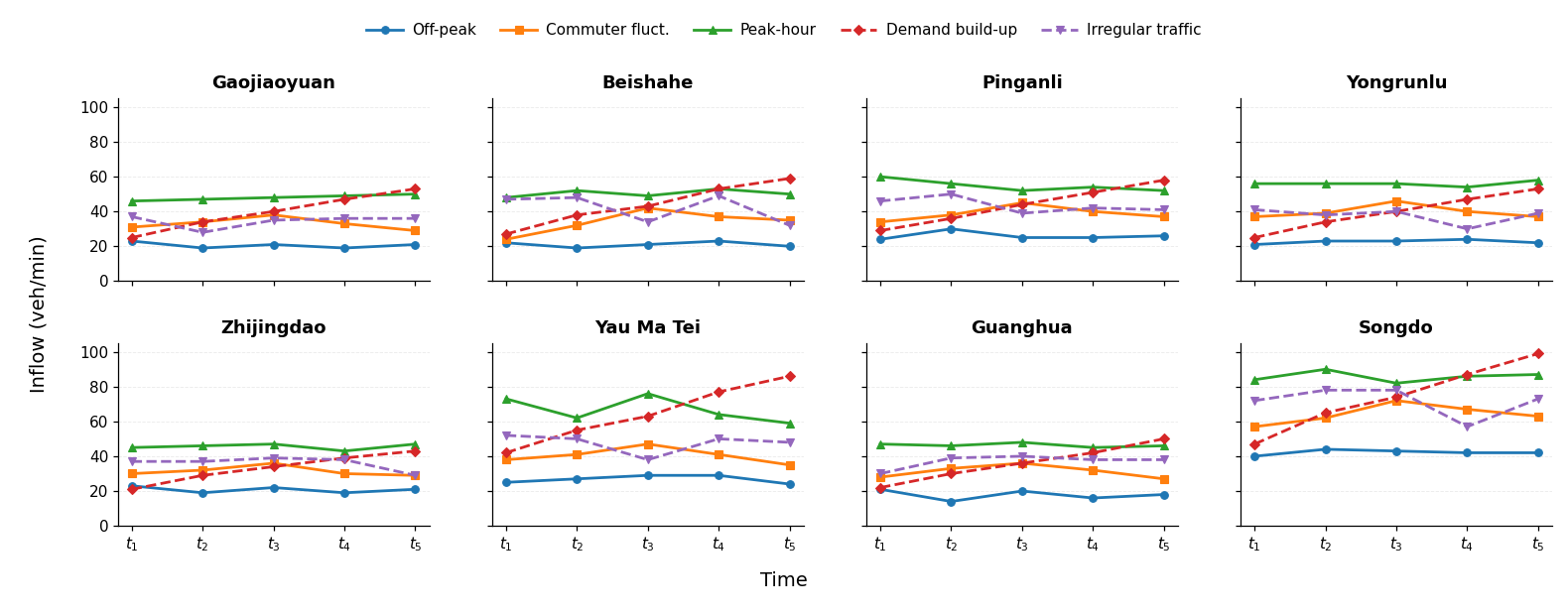}
    \caption{Traffic-demand profiles for the eight simulated scenes.}
    \label{fig:traffic_demand_profiles}
\end{figure*}

\begin{table*}[t]
    \centering
    \caption{Traffic-demand statistics across the eight simulated scenes. Each entry reports the mean $\pm$ standard deviation of the vehicle inflow over five time intervals, in vehicles per minute.}
    \label{tab:traffic_demand_stats}
    \small
    \setlength{\tabcolsep}{4pt}
    \begin{tabular}{@{}lccccc@{}}
        \toprule
        & \multicolumn{3}{c}{Training} & \multicolumn{2}{c}{Held-out Evaluation} \\
        \cmidrule(lr){2-4}
        \cmidrule(lr){5-6}
        Scene
        & Off-peak
        & \shortstack{Commuter fluctuation}
        & Peak-hour
        & \shortstack{Demand build-up}
        & \shortstack{Irregular traffic} \\
        \midrule
        Gaojiaoyuan & $20.6 \pm 1.5$ & $33.0 \pm 3.0$ & $48.0 \pm 1.4$ & $39.8 \pm 9.8$ & $34.4 \pm 3.3$ \\
        Beishahe & $21.0 \pm 1.4$ & $34.0 \pm 6.0$ & $50.4 \pm 1.9$ & $44.0 \pm 11.2$ & $42.0 \pm 7.4$ \\
        Pinganli & $26.0 \pm 2.1$ & $38.8 \pm 3.7$ & $54.8 \pm 3.0$ & $43.6 \pm 10.3$ & $43.6 \pm 3.9$ \\
        Yongrunlu & $22.6 \pm 1.0$ & $39.8 \pm 3.3$ & $56.0 \pm 1.3$ & $39.8 \pm 9.8$ & $37.6 \pm 3.9$ \\
        Zhijingdao & $20.8 \pm 1.6$ & $31.4 \pm 2.5$ & $45.6 \pm 1.5$ & $33.2 \pm 7.7$ & $36.0 \pm 3.6$ \\
        Yau Ma Tei & $26.8 \pm 2.0$ & $40.4 \pm 4.0$ & $66.8 \pm 6.6$ & $64.6 \pm 15.6$ & $47.6 \pm 5.0$ \\
        Guanghua & $17.8 \pm 2.6$ & $31.2 \pm 3.3$ & $46.4 \pm 1.0$ & $36.0 \pm 9.6$ & $37.0 \pm 3.6$ \\
        Songdo & $42.2 \pm 1.3$ & $64.2 \pm 5.0$ & $85.8 \pm 2.7$ & $74.4 \pm 17.9$ & $71.6 \pm 7.7$ \\
        \bottomrule
    \end{tabular}
\end{table*}

\subsection{Traffic Demand}
\label{app:traffic_demand}

For each scene, we construct five demand patterns with different traffic patterns. Three patterns are used for training: \emph{off-peak}, characterized by low and relatively stable inflow; \emph{peak-hour}, characterized by sustained high inflow; and \emph{commuter fluctuation}, which introduces structured variations over time. For evaluation, we introduce two held-out demand patterns that are not observed during training. \emph{Demand build-up} progressively increases the inflow over time, while \emph{irregular traffic} introduces stochastic temporal variations. 

\Figref{fig:traffic_demand_profiles} shows the temporal profiles of the five demand patterns across the eight scenes. Although the absolute inflow varies across scenes, the demand patterns preserve their intended temporal characteristics. Table~\ref{tab:traffic_demand_stats} further summarizes the mean and standard deviation of each demand profile over the five time intervals.

\subsection{Real-UAV Data}
\label{app:real_uav_data}

For real-world evaluation, we use UAV videos from the dataset introduced by Fonod et al.~\cite{fonod2025advanced}, selecting four intersections in Songdo, denoted as O, P, T, and H. The videos are captured at approximately 140--150\,m above ground and provide bird's-eye views of diverse intersection geometries and lane configurations. These scenes are used to evaluate the transfer of BEVLight from simulation to real UAV observations.

\begin{figure}[t]
    \centering
    \includegraphics[width=\textwidth]{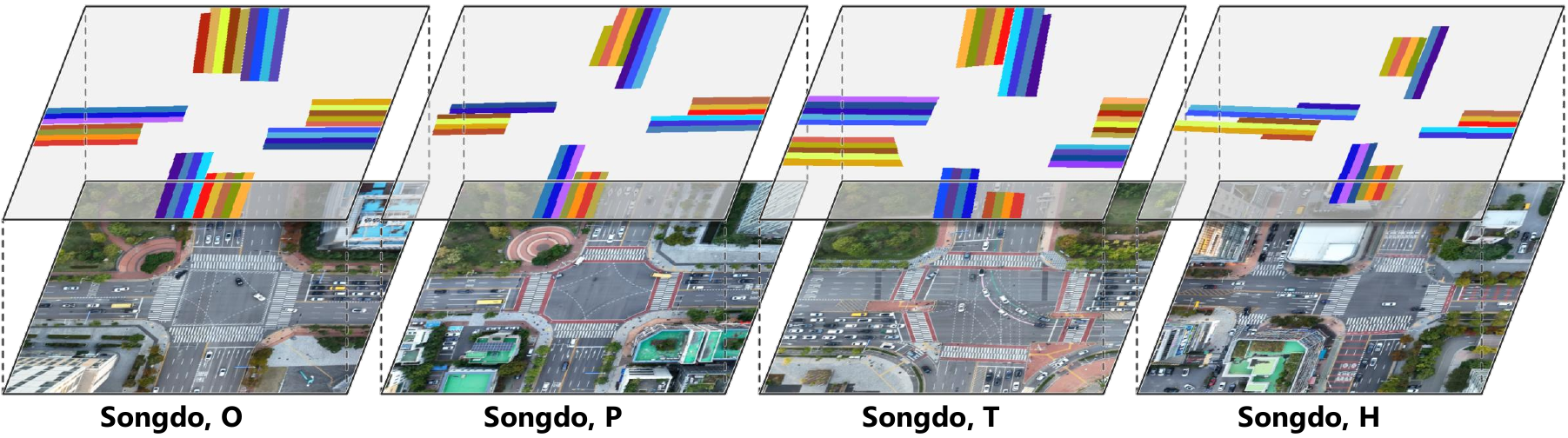}
    \caption{Real UAV observations and corresponding lane masks for intersections O, P, T, and H.}
    \label{fig:appendix_real_lane_mask}
\end{figure}

\Figref{fig:appendix_real_lane_mask} shows representative UAV observations and the corresponding lane masks for the four intersections. The masks indicate the visible lane regions used by BEVLight and illustrate how the lane-level representation is transferred from simulation to real UAV imagery. Table~\ref{tab:real_uav_lane_length} further reports the visible lane lengths in each scene.

\begin{table}[t]
    \centering
    \caption{Spatial coverage of the four real-UAV intersections.}
    \label{tab:real_uav_lane_length}
    \small
    \setlength{\tabcolsep}{5pt}
    \begin{tabular}{@{}lccccc@{}}
        \toprule
        & \multicolumn{2}{c}{Ground coverage}
        & \multicolumn{3}{c}{Visible lane length (m)} \\
        \cmidrule(lr){2-3}
        \cmidrule(lr){4-6}
        Intersection
        & Footprint (m $\times$ m)
        & Area (m$^2$)
        & Min.
        & Mean
        & Max. \\
        \midrule
        Songdo, O & $118.5 \times 118.5$ & 14,050 & 25.61 & 30.85 & 37.90 \\
        Songdo, P & $119.3 \times 119.3$ & 14,239 & 22.28 & 27.11 & 32.90 \\
        Songdo, T & $120.7 \times 120.7$ & 14,568 & 14.53 & 25.53 & 43.24 \\
        Songdo, H & $121.2 \times 121.2$ & 14,695 & 17.54 & 25.06 & 51.38 \\
        \bottomrule
    \end{tabular}
\end{table}

\subsection{Compared Methods}
\label{app:compared_methods}

We compare BEVLight with three groups of traffic signal methods: two conventional methods, four structured-state RL methods, and two vision-based two-stage methods. The conventional and RL methods directly access structured traffic states provided by the simulator, whereas the vision-based methods first extract lane-level traffic information from BEV observations before applying an RL controller.

\paragraph{Conventional Controllers.}
We include two commonly used traffic-responsive controllers:
\begin{itemize}[leftmargin=*]
    \item \textbf{FixedTime} follows a predefined signal schedule and holds each phase for 20\,s in our experiment, independent of the observed traffic conditions.
    \item \textbf{MaxPressure}~\citep{varaiya2013max} selects the signal phase with the highest traffic pressure, computed from the imbalance between upstream and downstream traffic states. It provides a strong adaptive baseline without requiring policy learning.
\end{itemize}

\paragraph{Structured-State RL Methods.}
These methods receive lane-level traffic states directly from the simulator and learn signal-control policies through RL:
\begin{itemize}[leftmargin=*]
    \item \textbf{UniTSA}~\citep{wang2024unitsa} introduces a junction-matrix representation to accommodate different intersection layouts and uses traffic-state augmentation to expose the policy to diverse intersection structures and traffic conditions.
    \item \textbf{IntelliLight}~\citep{wei2018intellilight} learns adaptive signal control from structured intersection states using a value-based RL formulation.
    \item \textbf{AttendLight}~\citep{oroojlooy2020attendlight} employs attention mechanisms to aggregate traffic information across lanes and candidate signal phases, allowing a shared policy to operate across different intersection configurations.
    \item \textbf{MPLight}~\citep{chen2020toward} incorporates traffic-pressure information into RL-based signal control and models competition among candidate phases.
\end{itemize}

\paragraph{Vision-Based Two-Stage Methods.}
We further construct two baselines that first recover traffic states from BEV observations and then use AttendLight for signal control:
\begin{itemize}[leftmargin=*]
    \item \textbf{YOLO-Count} uses YOLOv11~\cite{khanam2024yolov11} to detect vehicles from BEV images with oriented bounding boxes and assigns the detections to individual lanes to obtain lane-level vehicle counts. The detector is fine-tuned on synthetic images generated from the same training scenes used by BEVLight, ensuring a fair comparison without introducing additional visual training data. The resulting lane-level states are then provided to AttendLight for phase selection.
    \item \textbf{BEVLight-State+AttendLight} uses BEVLight's auxiliary lane-level prediction heads to estimate traffic states directly from BEV observations. These predicted states are subsequently passed to AttendLight for signal control, providing a two-stage alternative that does not rely on explicit object detection.
\end{itemize}

In contrast to the two vision-based baselines, BEVLight directly maps BEV observations and junction topology to signal actions without an intermediate vehicle-detection or structured-state reconstruction stage.

\section{Human Demonstrations}
\label{app:human_demonstrations}

\subsection{Web-Based Human Demonstration Collection}
\label{app:human_demo_collection}

\begin{figure}[t]
    \centering
    \includegraphics[width=0.95\linewidth]{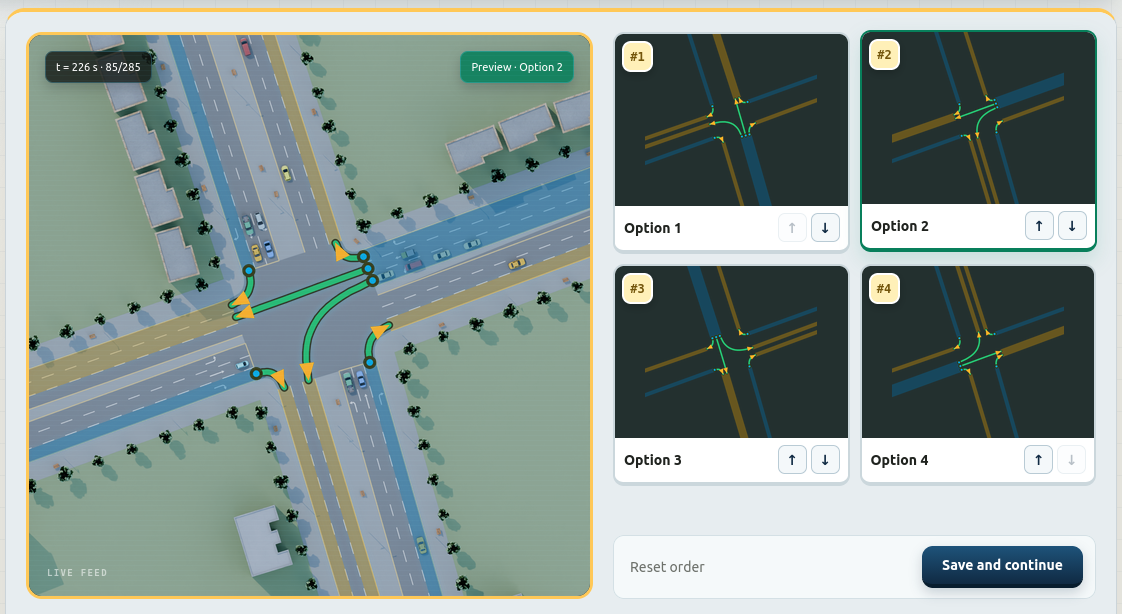}
    \caption{
    Web-based virtual TSC platform for collecting human demonstrations. The interface displays the current bird's-eye-view traffic scene and all feasible signal actions. Hovering over an action highlights its corresponding lanes and permitted movements on the intersection.
    }
    \label{fig:human_demo_interface}
\end{figure}

We develop a virtual teleoperation platform to collect human traffic-control demonstrations. Eight graduate researchers with backgrounds in computer science or transportation, including master's and PhD students, participated in the data collection. As illustrated in Fig.~\ref{fig:human_demo_interface}, the interface presents a bird's-eye view of the current intersection together with all feasible signal actions. To make the control semantics of each action explicit, hovering over an action highlights the corresponding lanes and permitted traffic movements directly on the intersection view.

Participants make a control decision every 10~s and rank all feasible actions using arrow-based controls. Rather than recording only the selected action, the ranking provides relative preference supervision among feasible signal phases, indicating not only which action is preferred but also how the remaining alternatives are ordered. This provides richer supervision for learning the phase-scoring function and avoids treating all non-selected actions as equally undesirable. For each decision, we record the complete simulator state, including the positions of all vehicles, together with the human-provided action ranking. The recorded vehicle states are subsequently used to reconstruct the same traffic scene under different visual appearances, as described in Appendix~\ref{app:human_demo_rendering}.

To ensure the quality of the demonstrations, we use MaxPressure as a reference controller and perform filtering at the episode level. Each human-controlled episode is compared with the MaxPressure controller under the same traffic scenario, and only episodes achieving a lower average waiting time are retained. After filtering, the dataset contains more than 5K human decisions, corresponding to approximately 14 hours of traffic-control demonstrations.

\subsection{Appearance-Diverse Scene Rendering}
\label{app:human_demo_rendering}

\begin{figure*}[p]
    \centering
    \includegraphics[width=0.9\textwidth]{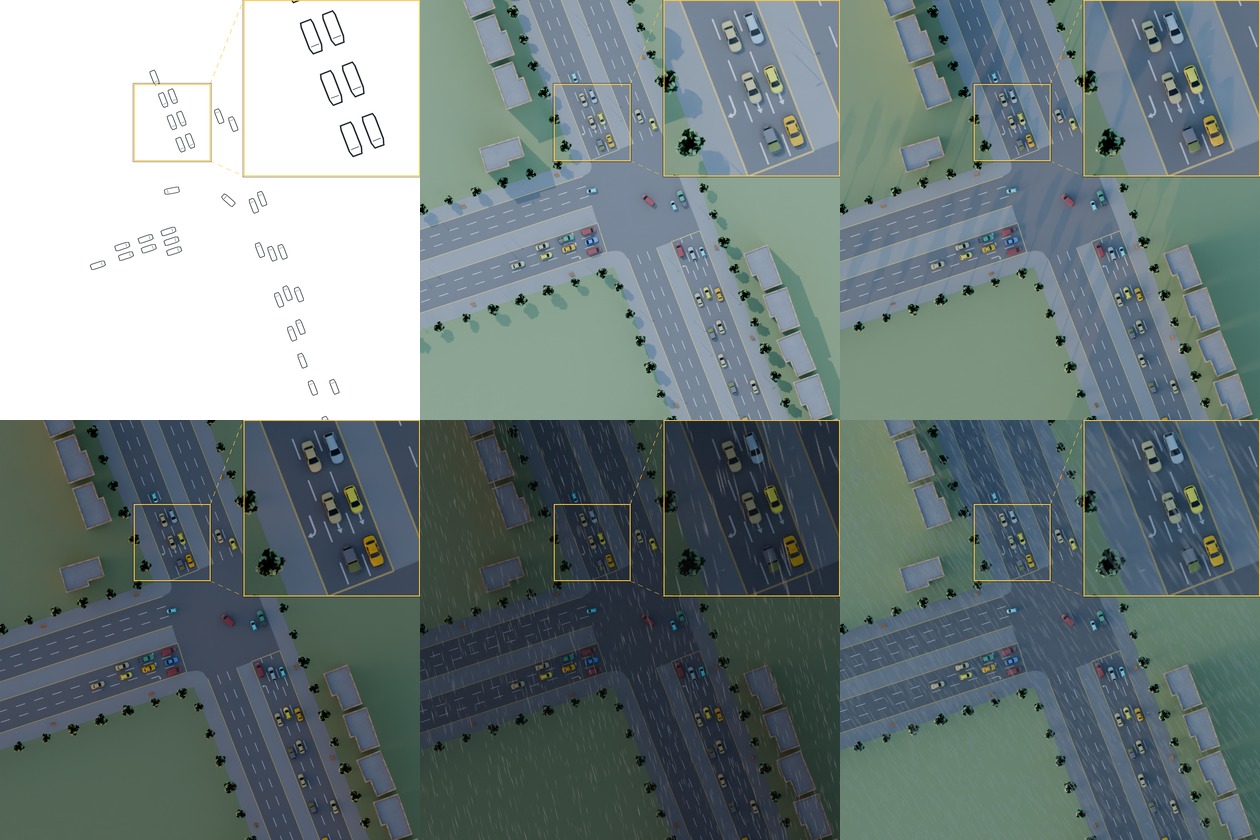}
    \par\vspace{0.4em}
    \textbf{(a) Beishahe, Beijing}
    
    \vspace{0.8em}
    
    \includegraphics[width=0.9\textwidth]{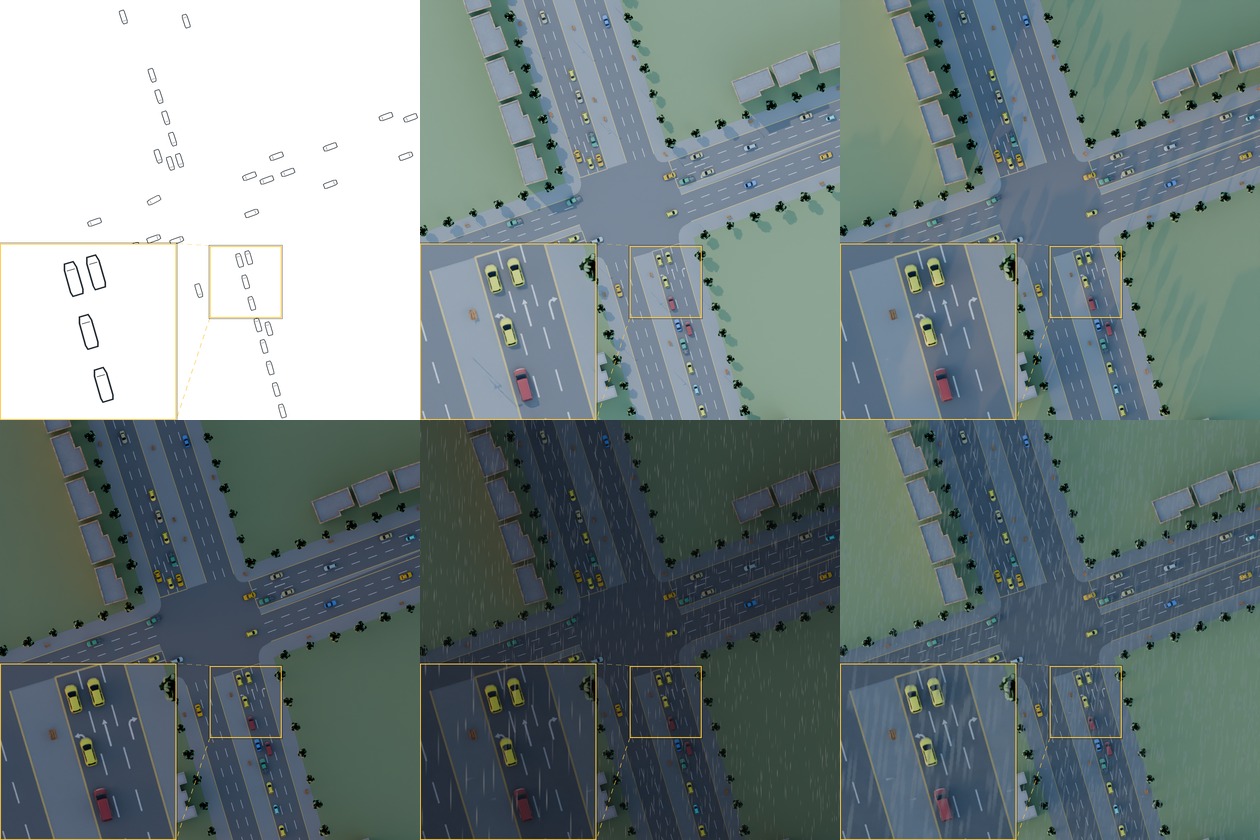}
    \par\vspace{0.4em}
    \textbf{(b) Gaojiaoyuan, Beijing}
    \caption{
    Appearance-diverse rendering of recorded human demonstration states. For each intersection, the top-left panel visualizes the recorded vehicle configuration, while the remaining panels show the same traffic state under five visual conditions: \textit{Morning}, \textit{Noon}, \textit{Dusk}, \textit{Dusk-Rain}, and \textit{Afternoon-Rain}. The underlying traffic state and human action ranking remain unchanged across all rendered variants. The examples shown here correspond to \textit{Beishahe, Beijing} and \textit{Gaojiaoyuan, Beijing}.
    }
    \label{fig:human_demo_rendering_1}
\end{figure*}

\begin{figure*}[p]
    \centering
    \includegraphics[width=0.9\textwidth]{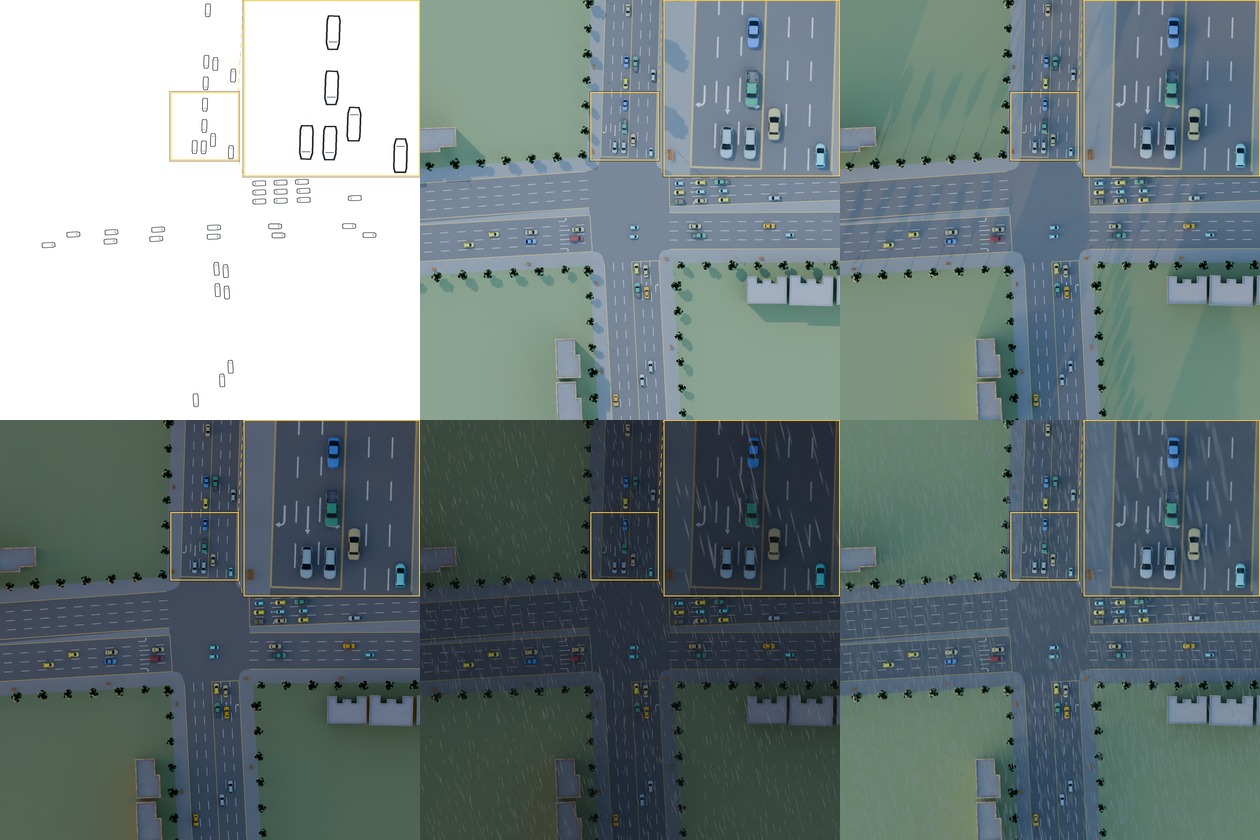}
    \par\vspace{0.4em}
    \textbf{(c) Pinganli, Beijing}

    \vspace{0.8em}
    \includegraphics[width=0.9\textwidth]{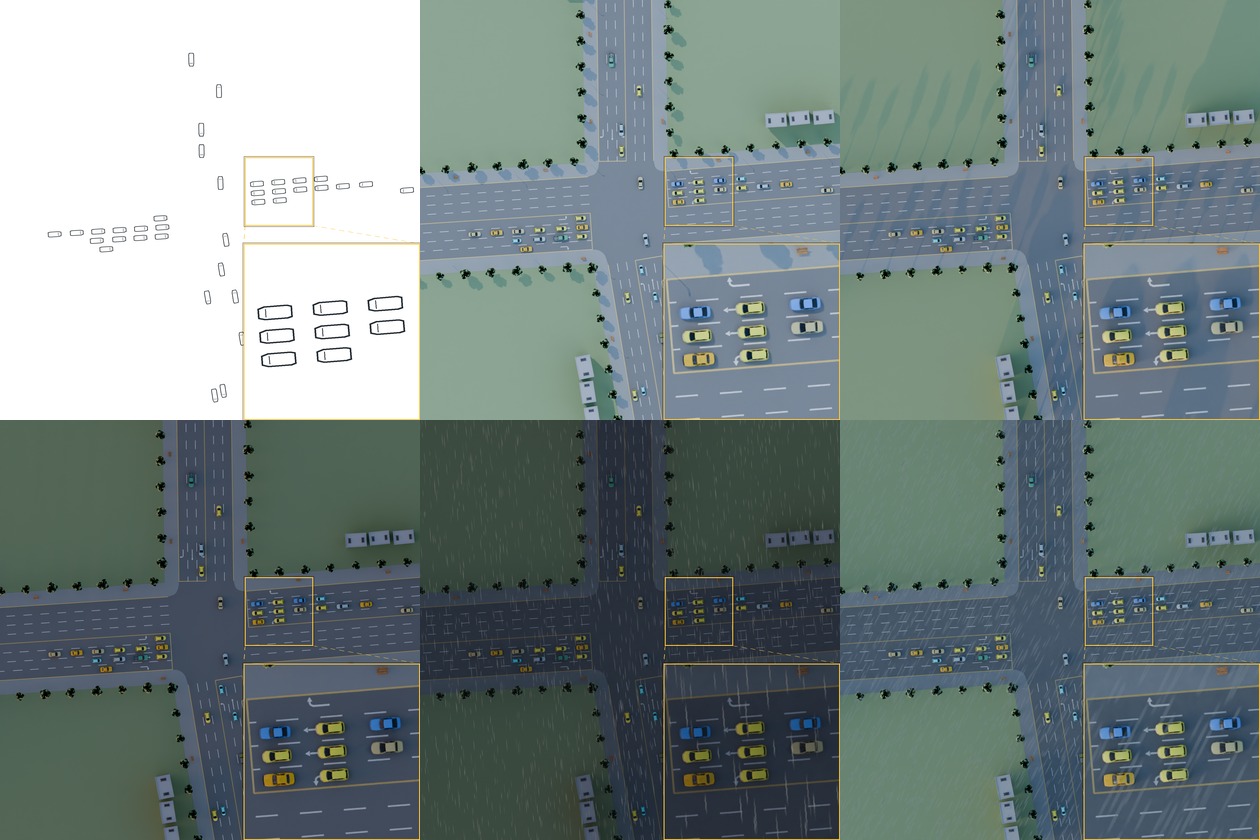}
    \par\vspace{0.4em}
    \textbf{(d) Yongrunlu, Beijing}

    \caption{
    Appearance-diverse rendering examples for
    \textit{Pinganli, Beijing} and \textit{Yongrunlu, Beijing}.
    }
    \label{fig:human_demo_rendering_2}
\end{figure*}

\begin{figure*}[p]
    \centering
    \includegraphics[width=0.9\textwidth]{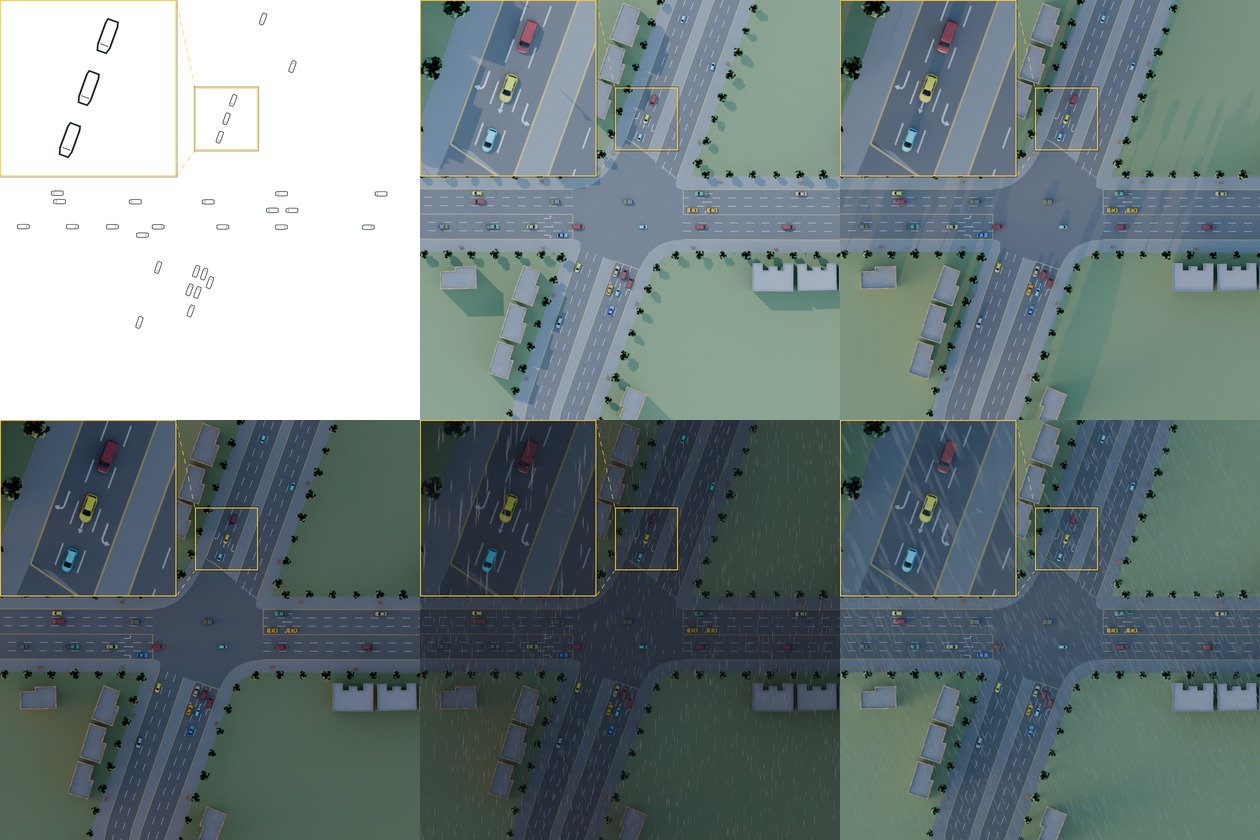}
    \par\vspace{0.4em}
    \textbf{(e) Guanghua, Chengdu}

    \vspace{0.8em}
    
    \includegraphics[width=0.9\textwidth]{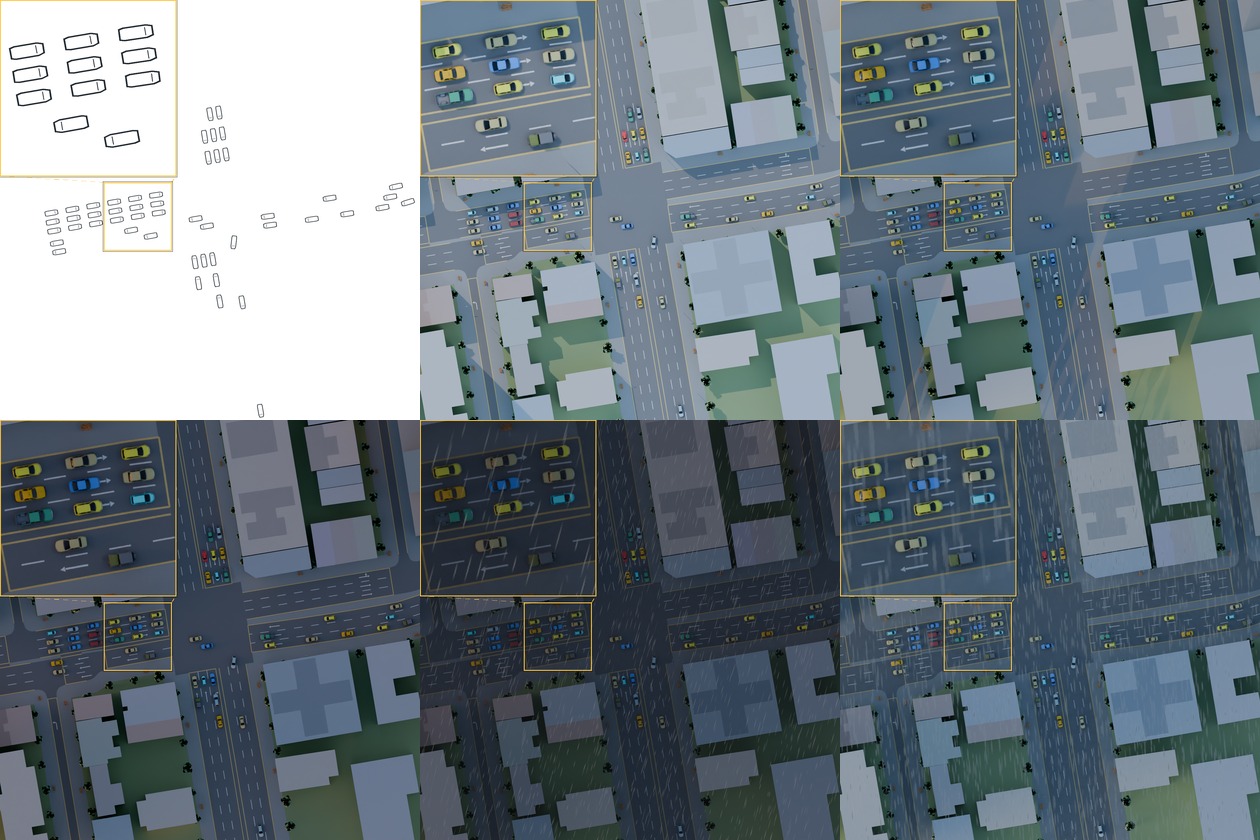}
    \par\vspace{0.4em}
    \textbf{(f) Yau Ma Tei, Hongkong}

    \caption{
    Appearance-diverse rendering examples for
    \textit{Guanghua, Chengdu} and \textit{Yau Ma Tei, Hongkong}.
    }
    \label{fig:human_demo_rendering_3}
\end{figure*}

\begin{figure*}[p]
    \centering
    \includegraphics[width=0.9\textwidth]{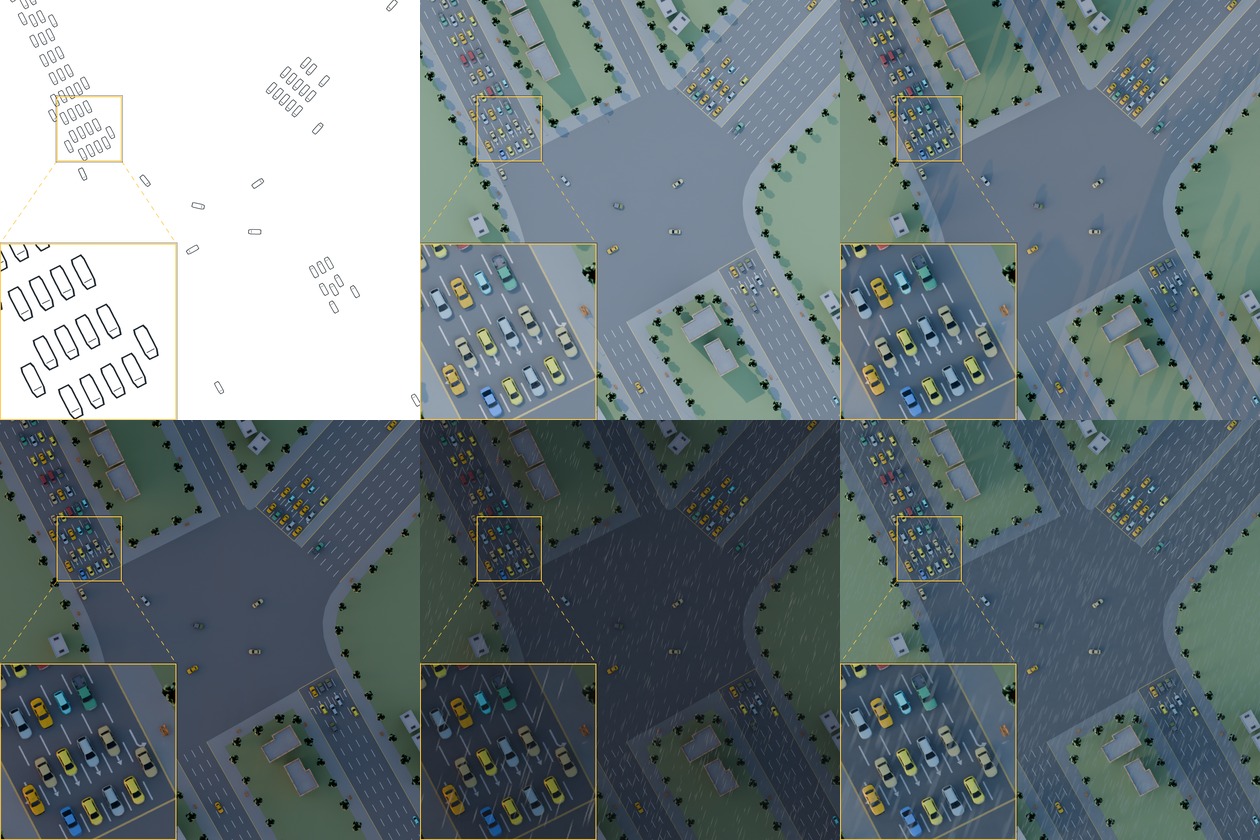}
    \par\vspace{0.4em}
    \textbf{(g) Songdo, SouthKorea}

    \vspace{0.8em}

    \includegraphics[width=0.9\textwidth]{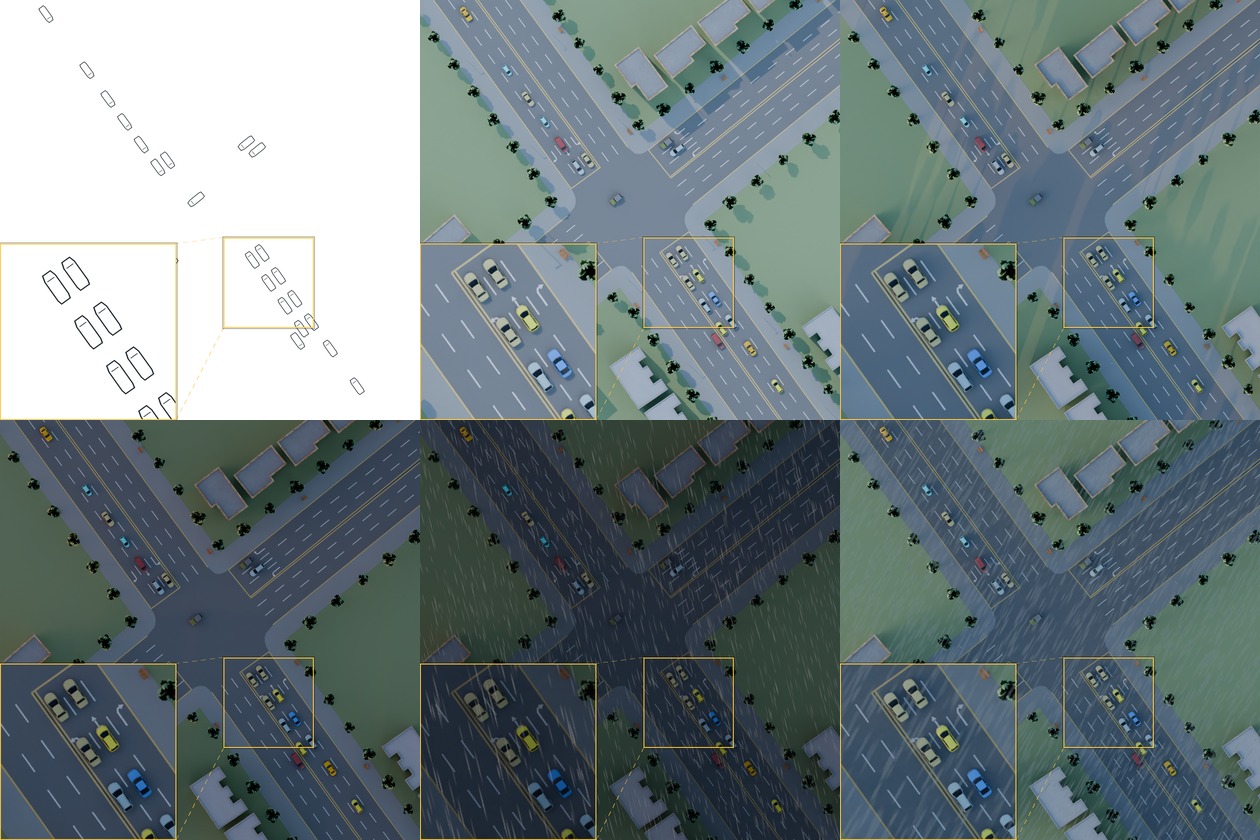}
    \par\vspace{0.4em}
    \textbf{(h) Zhijingdao, Tianjin}

    \caption{
    Appearance-diverse rendering examples for
    \textit{Songdo, SouthKorea} and \textit{zhijingdao, Tianjin}.
    }
    \label{fig:human_demo_rendering_4}
\end{figure*}

For each human decision, we record the underlying simulator state rather than storing only the rendered observation. In particular, the recorded state contains the positions of all vehicles at the decision time, allowing the same traffic configuration to be reconstructed independently of its original visual appearance. The associated human action ranking is preserved throughout this process. In this way, a single human demonstration can provide supervision for multiple visually different observations while retaining exactly the same traffic state and control preference.

We render each recorded state under five visual conditions: \textit{Morning}, \textit{Noon}, \textit{Dusk}, \textit{Dusk-Rain}, and \textit{Afternoon-Rain}. The rendering process changes illumination, weather, and scene appearance while keeping vehicle positions, junction geometry, and control supervision unchanged. As illustrated in Figs.~\ref{fig:human_demo_rendering_1}--\ref{fig:human_demo_rendering_4}, each example additionally includes a visualization of the recorded vehicle configuration before rendering. The figures cover all eight intersections used in our experiments: \textit{Beishahe, Beijing}, \textit{Gaojiaoyuan, Beijing}, \textit{Pinganli, Beijing}, \textit{Yongrunlu, Beijing}, \textit{Guanghua, Chengdu}, \textit{Yau Ma Tai, Hongkong}, \textit{Songdo, SouthKorea}, and \textit{Zhijingdao, Tianjin}.

This state-preserving rendering strategy increases the visual diversity associated with the collected human demonstrations. More importantly, because the same control preference is observed across different appearances of the same traffic state, the model is encouraged to rely on traffic-relevant visual information rather than appearance-specific cues. We therefore use these appearance-diverse renderings throughout policy training as a component for improving robustness to visual-domain changes and supporting transfer from simulated BEV observations to real UAV imagery.

\begin{figure*}[t]
    \centering

    \includegraphics[width=0.99\textwidth]{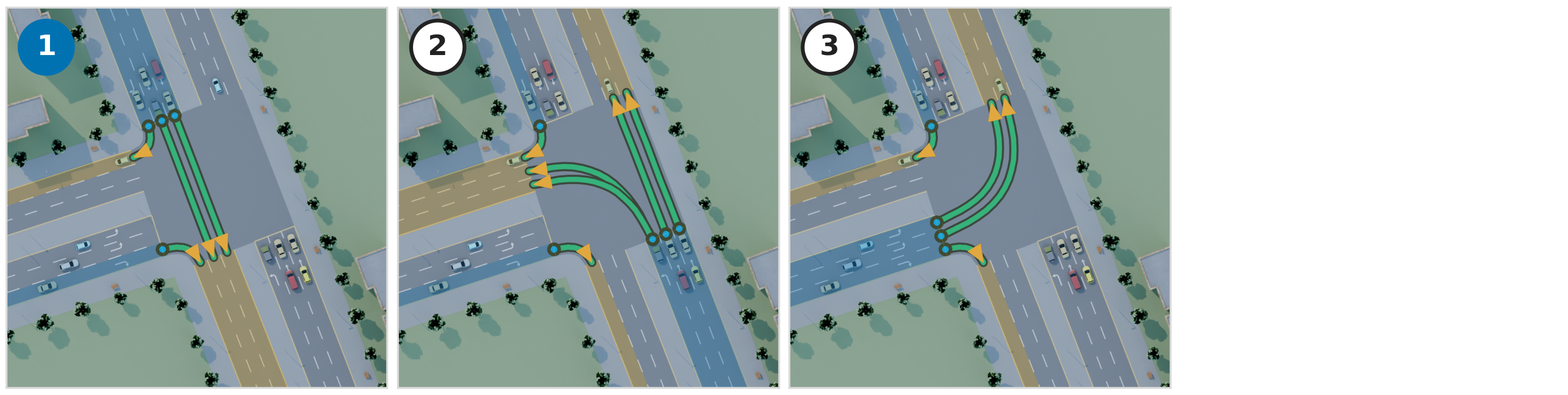}

    \vspace{0.2em}
    \textit{(a) Beishahe, Beijing}

    \vspace{0.8em}

    \includegraphics[width=0.99\textwidth]{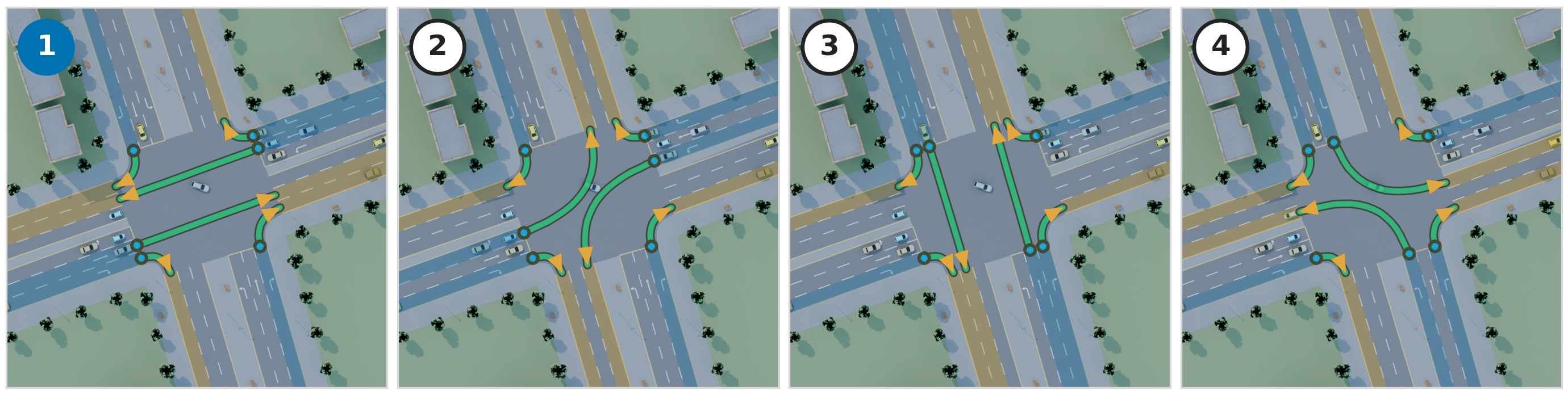}

    \vspace{0.2em}
    \textit{(b) Gaojiaoyuan, Beijing}

    \vspace{0.8em}

    \includegraphics[width=0.99\textwidth]{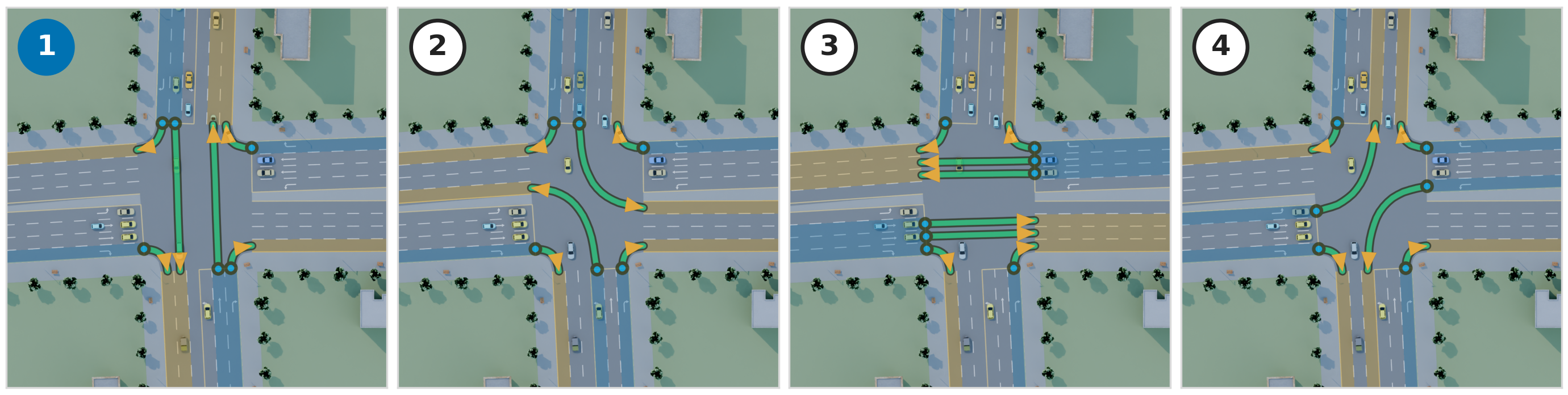}

    \vspace{0.2em}
    \textit{(c) Pinganli, Beijing}

    \vspace{0.8em}

    \includegraphics[width=0.99\textwidth]{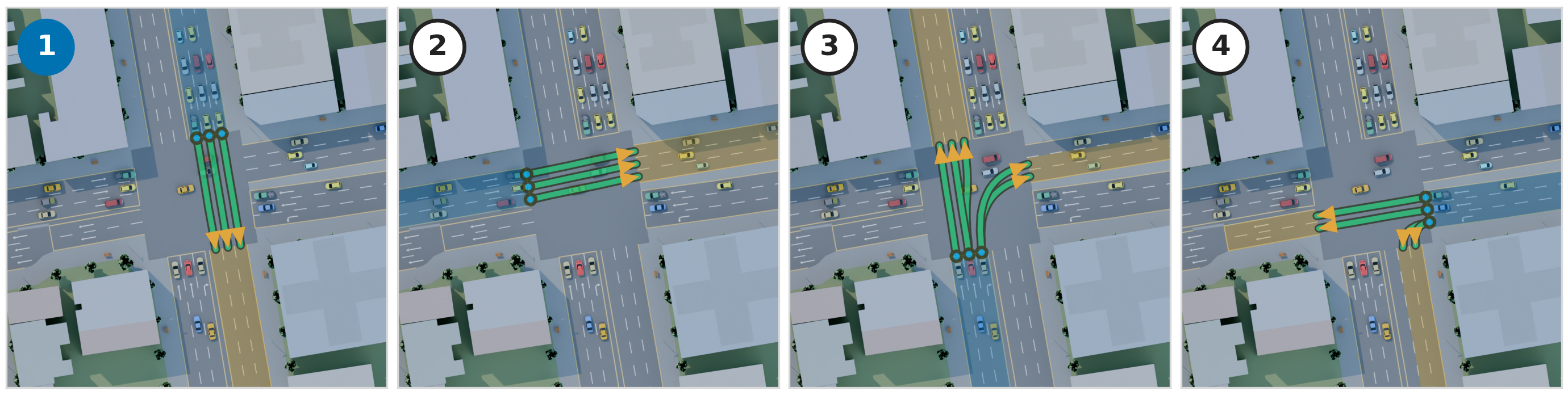}

    \vspace{0.2em}
    \textit{(d) Yau Ma Tei, Hong Kong}

    \caption{
    Representative examples of human control preferences and action rankings.
    Each example shows a traffic observation together with the ranking assigned by the human operator over all feasible signal phases.
    The ranking preserves the relative ordering among feasible actions rather than retaining only the top-ranked phase, providing richer supervision for learning the phase-scoring function in BEVLight.
    }
    \label{fig:human_ranking_cases}
\end{figure*}

\subsection{Human Control Preferences and Action Ranking}
\label{app:human_action_ranking}

In addition to the selected action, we collect a complete ranking over all feasible signal phases at each decision step. The motivation is that a single selected-action label provides only the top choice, while all non-selected actions would otherwise be treated equivalently. In the TSC task, however, feasible actions that are not selected can still differ substantially in their relative priority. By recording the full ranking, we preserve this relative preference structure and provide richer supervision for learning the phase-scoring function in BEVLight.

Fig.~\ref{fig:human_ranking_cases} presents representative examples from four intersections: \textit{Beishahe, Beijing}, \textit{Gaojiaoyuan, Beijing}, \textit{Pinganli, Beijing}, and \textit{Yau Ma Tei, Hong Kong}. Each example shows the traffic observation at a decision step together with the human-assigned ranking over all feasible signal phases.

More specifically, the ranking supervision specifies not only the preferred phase, but also the relative ordering of the remaining feasible phases. In \Figref{fig:human_ranking_cases}(a), the action ranked third serves no vehicles under the current traffic state, whereas the second-ranked action serves active traffic, providing a meaningful distinction between two non-selected actions. Meanwhile, the first- and second-ranked actions correspond to traffic conditions of comparable congestion and are both placed ahead of the remaining alternatives. In \Figref{fig:human_ranking_cases}(c), the ranking further reflects the current signal context: the active phase is prioritized to allow vehicles already being served to clear the intersection before switching to another phase. In \Figref{fig:human_ranking_cases}(d), the highest-ranked action serves visibly heavier traffic than the other alternatives and is accordingly assigned the highest priority. These examples illustrate that the full ranking preserves distinctions among both selected and non-selected actions that would be lost if only the final action were recorded.

\section{Additional Real-World Case Studies}
\label{app:real_actions}

\begin{figure}[!t]
  \centering
  \includegraphics[width=\textwidth]{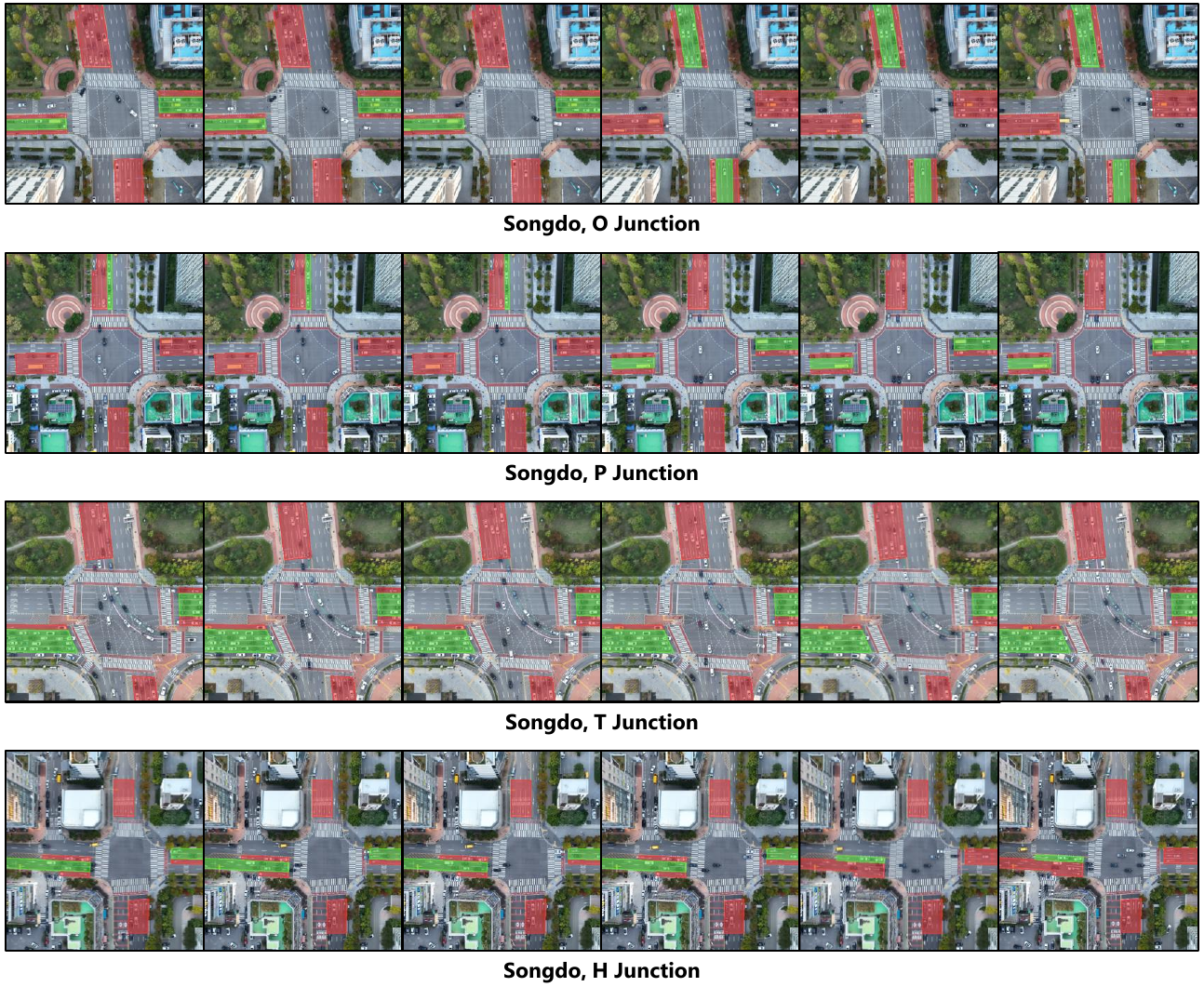}
  \caption{Zero-shot phase predictions over six consecutive UAV frames at Junctions O, P, T, and H in Songdo. Predictions are evaluated offline without controlling actual signals.}
  \label{fig:appendix_real_actions}
\end{figure}

\Figref{fig:appendix_real_actions} presents six consecutive UAV frames from each of four real intersections in Songdo: Junctions O, P, T, and H, together with BEVLight's predicted signal phases. The model is applied directly to real UAV imagery without real-world fine-tuning. Across these sequences, its phase predictions respond to changes in vehicle distribution, shifting priority as traffic clears and demand changes across movements.

Since the evaluation uses recorded videos, the predicted phases do not affect actual signals or subsequent traffic. These examples therefore provide qualitative evidence of zero-shot transfer rather than closed-loop control performance: BEVLight produces plausible phase decisions that adapt to evolving traffic conditions under real-world visual inputs.

\section{Discussion}
\label{app:discussion}

\paragraph{Practical considerations.}
Deployment hinges on the two properties evaluated in Sec.~\ref{sec:experiments}: a policy that is not tied to a particular junction or signal plan can be moved between sites without retraining, and a policy trained entirely in simulation can be applied to real UAV observations without collecting and labeling real control data, which is impractical at scale. Beyond the UAV video stream, BEVLight requires lane masks in the BEV frame and the lane-to-movement and movement-to-phase relations of the active signal plan. The relations are part of any signal timing plan, and the masks are specified once per junction and viewpoint. Decisions are made every 10\,s, leaving ample margin for transmission and inference. The policy only selects among feasible phases, so it cannot release conflicting movements; phase transitions and minimum green times remain under the signal controller, and a human operator can override the policy at any step. UAV observation is unavailable under strong wind, heavy precipitation, or restricted airspace, so BEVLight is best viewed as an on-demand controller that complements installed infrastructure.

\paragraph{Limitations.}
Our simulation-to-real evaluation is offline: queue estimates and phase decisions transfer to recorded UAV videos, but we do not control real signals, and at the two sites with the sparsest queues the queue-count error remains above an always-zero predictor. BEVLight controls a single junction, and coordination across junctions is left for future work. The policy imitates demonstrations from graduate researchers filtered against MaxPressure, so it inherits their quality and is not optimized for a long-horizon objective. Lane masks are assumed static, so UAV drift would require image registration. Finally, the simulated UAV height of 100\,m differs from the 140--150\,m of the real recordings, and appearance rendering covers daytime and rain but not night-time.

\paragraph{Broader impact.}
BEVLight lowers the infrastructure barrier for adaptive signal control and can be deployed on demand where fixed sensing is absent or temporarily disrupted, with reduced waiting time translating into lower fuel consumption and emissions. UAV imagery raises privacy and regulatory concerns: BEVLight operates on lane-level features from images captured above 100\,m, at which individuals and license plates are not identifiable, and does not track vehicles across frames, but deployments should follow local airspace regulations and data-retention policies. The demonstrations were collected in simulation from consenting participants and contain no personal data. Since a learned controller reflects the priorities of its demonstrations, operators should verify that its phase preferences do not systematically disadvantage particular approaches before deployment. The virtual teleoperation platform is not specific to BEVLight. It records complete simulator states alongside human decisions, so any recorded demonstration can be re-rendered under new appearances or used to supervise other visual controllers, and it can be extended to other observation modalities such as roadside cameras. We plan to release the platform and the collected demonstrations to support future work on human-supervised visual TSC.

\end{document}